\documentclass[12pt]{article}
\usepackage[T1]{fontenc}
\usepackage[utf8x]{inputenc}
\usepackage{graphicx}

\usepackage{todonotes}
\usepackage{rotating}
\usepackage{multirow}
\usepackage{nicefrac}
\usepackage{tabularx}
\usepackage{tabulary}
\usepackage{afterpage}
\usepackage{placeins}
\usepackage{booktabs}
\usepackage{setspace}
\usepackage{color}
\usepackage{setspace}
\usepackage{bigstrut}
\usepackage[bottom]{footmisc}
\usepackage[colorlinks=true,urlcolor=blue,linkcolor=blue,citecolor=blue]{hyperref}
\usepackage{threeparttable}
\usepackage{dcolumn}
\newcolumntype{G}{D..{6.4}}
\usepackage{amsmath}
\usepackage{amssymb}
\usepackage{enumerate}   
\usepackage{array}
\usepackage{pdflscape}
\newcolumntype{L}[1]{>{\raggedright\let\newline\\\arraybackslash\hspace{0pt}}m{#1}}
\newcolumntype{C}[1]{>{\centering\let\newline\\\arraybackslash\hspace{0pt}}m{#1}}
\newcolumntype{R}[1]{>{\raggedleft\let\newline\\\arraybackslash\hspace{0pt}}m{#1}}
\newcolumntype{Y}{>{\centering\arraybackslash}X}

\usepackage{siunitx}
\usepackage{newtxtext,newtxmath}

\usepackage{eqparbox}

\usepackage{rotating}

\usepackage{caption}
\usepackage{subcaption}
\makeatletter\let\p@subfigure\thefigure\makeatother

\newcommand{\blind}{0}

\usepackage{natbib}
\let\natbibcitet\citet
\renewcommand\citet{\bibpunct{(}{)}{,}{a}{,}{,}\natbibcitet}
\let\natbibcitep\citep
\renewcommand\citep{\bibpunct{(}{)}{;}{a}{,}{;}\natbibcitep}
\newcommand{\bi}{\begin{itemize}}
\newcommand{\ei}{\end{itemize}}
\newcommand{\be}{\begin{equation}}
\newcommand{\ee}{\end{equation}}
\defcitealias{ieo14}{IEO, 2014}

\begin{document}

\def\spacingset#1{\renewcommand{\baselinestretch}%
{#1}\small\normalsize} \spacingset{1}


\if0\blind
{
  \title{\bf A Bayesian approach for estimation of weight matrices in spatial autoregressive models\thanks{This working paper is an earlier draft of an article published by Taylor \& Francis in Spatial Economic Analysis on 22nd July 2022, available at: \url{https://www.tandfonline.com/doi/full/10.1080/17421772.2022.2095426}.}
  }
  \author{Tam\'{a}s Krisztin\thanks{Tam\'{a}s Krisztin was supported by funds of the Austrian National Bank: 18690.} \hspace{.2cm}\\
    International Institute for Applied Systems Analysis (IIASA)\\
    and \\
    Philipp Piribauer\thanks{Philipp Piribauer was supported by the Austrian Science Fund (FWF): ZK 35.}\\
    Austrian Institute of Economic Research (WIFO)}
  \maketitle
} \fi

\if1\blind
{
  \bigskip
  \bigskip
  \bigskip
  \begin{center}
    {\LARGE\bf A Bayesian approach for estimation of weight matrices in spatial autoregressive models}
\end{center}
  \medskip
} \fi

\bigskip
\begin{abstract}
\noindent We develop a Bayesian approach to estimate weight matrices in spatial autoregressive (or spatial lag) models. Datasets in regional economic literature are typically characterized by a limited number of time periods $T$ relative to spatial units $N$. When the spatial weight matrix is subject to estimation severe problems of over-parametrization are likely. To make estimation feasible, our approach focusses on spatial weight matrices which are binary prior to  row-standardization. We  discuss the use of hierarchical priors which impose sparsity in the spatial weight matrix. Monte Carlo simulations show that these priors perform very well where the number of unknown parameters is large relative to the observations. The virtues of our approach are demonstrated using global data from the early phase of the COVID-19 pandemic. %


\end{abstract}

\noindent%
\vfill

\newpage
\spacingset{1.45} 
\section{Introduction}

Spatial econometrics deals with the study of cross-sectional dependence and interactions among (spatial) observations. A particularly popular spatial econometric model is the spatial autoregressive (or spatial lag) specification, where spatial interdependence between observations  is governed by a so-called spatial weight matrix. The spatial weight matrix is typically assumed non-negative, row-standardized and exogenously given, with spatial weights based on some concept of neighbourhood. Geographic neighbourhood is often preferred due to exogeneity assumptions. However, when relying on geographic information, several competing approaches exist for constructing the weight matrix  (for a thorough discussion, see \citealt{LeSage2009}).  Recently, \cite{kelejian2014estimation}, \cite{qu2015estimating},  \cite{han2016bayesian}, and \cite{hsieh2016social} use alternative measures, such as (socio-)economic proximity. Another strand of the literature focusses on the uncertainty associated with the choice of neighbourhood structures by selecting or combining alternative weight matrices (see, for example, \citealt{debarsy2018flexible} and \citealt{Piribauer2015}).  

Since direct estimation of a spatial weight matrix requires estimating at least $(N-1)N$ parameters (ignoring the other model parameters), only few approaches target direct estimation of spatial weight matrices. Recently, \cite{ahrens2015two} and \cite{Lam2020a} tackle this problem through LASSO-based approaches (\citealt{tibshirani1996regression}), which involve (a priori) expert knowledge about the interactions between spatial units, while allowing the final estimates of the spatial weights to slightly deviate from it.\footnote{\cite{ahrens2015two} consider the case of sparsity in the spatial weights by employing shrinkage towards the zero matrix.}  However, for regional economic panels, where the time dimension $T$ is often limited relative to the number of spatial observations $N$,  estimation results in a deleterious proliferation of the number of parameters. 

In this paper we describe a novel and flexible Bayesian approach for estimation of spatial weight matrices. Our definition of spatial weight matrices fulfils the typical assumptions employed in the vast majority of spatial econometric literature. The resulting spatial weight matrices are assumed non-negative and specific requirements to identification of the parameters can be easily implemented in a Markov-chain Monte Carlo (MCMC) sampling strategy. Although our primary focus is on row-standardized spatial weight matrices, weights without row-standardization are also implementable. To make our estimation approach applicable to spatial panels where the number of time periods $T$ is limited as compared to the number of spatial units $N$, we focus on spatial weight matrices which are binary prior to potential row-standardization.

In this paper we primarily focus on scenarios where no a priori information on the spatial structure is available. However, we also discuss how a priori spatial information can be implemented in a very simple and transparent way. For cases where the number of unknown parameters is large relative to the number of observations, we discuss hierarchical prior setups which impose sparsity in the weight matrix. In a Monte Carlo study, we show that these sparsity priors perform particularly well when the number of spatial observations $N$ is large relative to the time periods $T$.

We show that our approach can be implemented in an efficient Gibbs sampling algorithm, which implies that the estimation strategy can be easily extended to other spatial econometric specifications. Among several others, such extensions include shrinkage estimation to avoid overparameterization (\citealt{Piribauer2015}), more flexible specifications of the innovation process (\citealt{lesage1997bayesian}), controlling for unobserved spatial heterogeneity (\citealt{cornwall2017embracing}; \citealt{piribauer2016heterogeneity}), or allowing for non-linearity in the slope parameters (\citealt{Basile2008}; \citealt{Krisztin2017}). It is moreover worth noting that the proposed approach can be easily adapted to matrix exponential spatial specifications (\citealt{lesage2007mess}), spatial error specifications (see, \citealt{LeSage2009}), or local spillover models (\citealt{HalleckVega2015}).

The rest of the paper is organized as follows: the next section outlines the panel version of the considered spatial lag model. Section 3 discusses the Bayesian estimation approach of the spatial weights along with several potential prior setups. Section 4 presents the Bayesian MCMC estimation algorithm and also discusses how to efficiently deal with the computational difficulties when updating the spatial weights in the MCMC sampler. Section 5 assesses the accuracy of the sampling procedure via a Monte Carlo simulation study. Section 6 illustrates our approach using data on global infection rates of the very first phase of the recent COVID-19 pandemic. The final section concludes.

\section{Econometric framework}
We consider a panel version of a global spillover spatial autoregressive model (SAR) of the form:\footnote{We also consider specifications with a spatial lag of the temporally lagged dependent variable. Sampling strategies for these cases are presented in the appendix.}
\begin{equation}
\boldsymbol{y}_t= \rho\boldsymbol{Wy}_t+\boldsymbol{\mu}+\tau_t+\boldsymbol{Z}_t\boldsymbol{\beta}_0+\boldsymbol{\varepsilon}_t, \hspace{2cm}t=1,...,T
\end{equation}
where $\boldsymbol{y}_t$ denotes an $N\times 1$ vector of observations on the dependent variable measured at period $t$. $\boldsymbol{\mu}$ and $\tau_t$ represent parameters associated with fixed effects for the $N$ spatial units and $T$ time periods, respectively. $\boldsymbol{Z}_t$ is an $N\times q_0$ full rank matrix of explanatory variables, with corresponding $q_0\times 1$ vector of slope parameters $\boldsymbol{\beta}_0$. $\boldsymbol{\varepsilon}_t$ is a standard $N\times 1$ disturbance term $\boldsymbol{\varepsilon}_t\sim \mathcal{N}(\boldsymbol{0},\sigma^2\boldsymbol{I}_N)$. 

The $N\times N$ matrix $\boldsymbol{W}$ denotes a spatial weight matrix and $\rho$ is a (scalar) spatial dependence parameter. $\boldsymbol{W}$ is non-negative with $w_{ij}>0$ if observation $j$ is considered as a neighbour to $i$, and $w_{ij}=0$ otherwise. A vital assumption is also that $w_{ii}=0$, in order to avoid the case that an observation is assumed as a neighbour to itself. A frequently made assumption amongst practitioners is that $\boldsymbol{W}$ is row-stochastic with rows summing to unity. In this paper, we mainly present results relating to row-stochastic weight matrices. However, as the decision on row-standardizing $\boldsymbol{W}$  depends on the empirical application, it is worth noting that the proposed approach may be easily adapted to problems without row-standardization of $\boldsymbol{W}$.\footnote{Thorough discussions on the implications of row-standardization are provided by \cite{plumper2010model} and \cite{liu2014endogenous}.}

The reduced form of the SAR model is given by:
\begin{equation}
\boldsymbol{y}_t= (\boldsymbol{I}_N-\rho\boldsymbol{W})^{-1}(\boldsymbol{\mu}+\tau_t+\boldsymbol{Z}_t\boldsymbol{\beta}_0+\boldsymbol{\varepsilon}_t),
\end{equation}
where $(\boldsymbol{I}_N-\rho\boldsymbol{W})^{-1}=\sum_{r=0}^\infty\rho^r\boldsymbol{W}^r$ is a so-called spatial multiplier matrix. To ensure that ($\boldsymbol{I}_N-\rho\boldsymbol{W}$) is invertible, appropriate stability conditions need to be imposed. For row-stochastic spatial weight matrices, a sufficient stability condition for the spatial autoregressive parameter often employed is $\rho\in(-1,1)$ (see, for example, \citealt{LeSage2009}).

In most cases, the elements of $\boldsymbol{W}$ are typically treated as known. In the spatial econometric literature, there are various ways as a means to constructing such a spatial weight matrix. In this study we focus on estimation of weight matrices which are binary prior to row-standardization. We therefore assume that the typical element of our spatial weight matrix can be obtained from an unknown $N\times N$ spatial adjacency matrix $\boldsymbol{\Omega}$ (with typical element $\omega_{ij}$).\footnote{Eq. (\ref{eq:rowstand}) implies some observations may have zero neighbours.  However, priors on the number of neighbours can be easily elicited to rule out such situations. Moreover, a researcher might easily abstain from row-standardization by neglecting the transformation in Eq. (\ref{eq:rowstand}).} We therefore define $\boldsymbol{W}=f(\boldsymbol{\Omega})$, where $f(\cdot)$ denotes the row-standardization function:\footnote{The function $f(\cdot)$ may simply be dropped when considering models without row-standardization of $\boldsymbol{W}$.}
\begin{equation}
\label{eq:rowstand}
w_{ij}=  \begin{cases}
\omega_{ij}/\sum_{j=1}^{N}\omega_{ij} & \text{if } \sum_{j=1}^{N}\omega_{ij} > 0\\
0 & \text{otherwise}.
\end{cases}
\end{equation}
The elements of the adjacency matrix $\boldsymbol{\Omega}$ are assumed as unknown binary indicators, which are subject to estimation. It is worth noting that the assumption of a binary $\boldsymbol{\Omega}$ covers a wide range of specifications commonly used in the literature such as contiguity, distance band, or nearest neighbours (see, for example, \citealt{LeSage2009}). 

To alleviate further notation, we collect the respective dummy variables associated with the fixed effects along with the explanatory variables in a $N\times q$ matrix $\boldsymbol{X}_t$ with corresponding $q\times 1$ parameter vector $\boldsymbol{\beta}$. Moreover, define $\boldsymbol{Y}=\left[\boldsymbol{y}_1', \dots,  \boldsymbol{y}_T' \right]'$, $\boldsymbol{X}=\left[\boldsymbol{X}_1', \dots, \boldsymbol{X}_T'\right]'$, $\boldsymbol{S}=  \boldsymbol{I}_{T} \otimes  (\boldsymbol{I}_{N}-\rho\boldsymbol{W})$, and $\mathcal{D}=\{\boldsymbol{Y},\boldsymbol{X}\}$ denotes the data. The Gaussian likelihood $p(\mathcal{D}|\bullet)$ is then given by:
\begin{equation}
p(\mathcal{D}|\bullet)= \frac{1}{(2\pi\sigma^2)^{NT}} |\boldsymbol{S}| \exp\left[-\frac{1}{2\sigma^2}(\boldsymbol{SY}-\boldsymbol{X\beta})'(\boldsymbol{SY}-\boldsymbol{X\beta})\right]. \label{eq:Likelihood}
\end{equation}
When the elements of the spatial weight matrix are subject to estimation, the number of unknown parameters is likely much larger than the number of observations. Since spatial economic panels often feature limited $T$ relative to $N$, the proposed estimation approach has to address the issue of over-parametrization. We  discuss different ways to tackle this problem. First and foremost, one may reduce the dimensionality of the problem by imposing a priori information on spatial weights or assuming symmetry of the spatial neighbourhood structure. Alternatively, we consider hierarchical prior setups which impose sparsity in the weight matrix.

When estimating spatial weights in addition to the spatial and slope parameters, identification issues are more complicated as compared to models assuming exogenous spatial weights. We therefore follow \cite{de2019identifying}, who provide a thorough discussion on parameter identification for rather general spatial autoregressive model specifications. As mentioned before, we consider spatial weight matrices which are non-negative and $w_{ii}=0$ for all $i$. Further standard assumptions include $\sum_{j=i}^{N}|\rho w_{ij}|<1$ $\forall i$, $|\rho| < 1$, and $||\boldsymbol{W}||<C$ for some positive $C\in \mathbb{R}$, as well as $\boldsymbol{\beta}_0\rho\neq 0$. As an additional identifying assumption, it is important that the main diagonal elements of $\boldsymbol{W}^2$ are not proportional to a vector of ones.\footnote{The most obvious case, where this assumption would be violated is a fully connected $\boldsymbol{W}$ with $w_{ij}=1/N$ for all $i\neq j$.} Sufficient conditions for global identification are fulfilled if we make the additional assumption of $\rho>0$ (see Corollary 3 in \citealt{de2019identifying}). Without this additional restriction on $\rho$, \cite{de2019identifying} show that a strongly connected spatial network for global identification is needed. Since strong a priori information on the spatial weight matrix is often not available (or desired), we therefore assume $\rho\in(0,1)$ and only consider positive spatial autocorrelation, which is a typical assumption for empirical applications.\footnote{These assumptions can be checked during estimation by using standard rejection sampling techniques in the MCMC sampling steps (see, for example, \citealt{LeSage2009}, or \citealt{Koop2003}). Rejection sampling rejects draws of parameter combinations which do not fulfil these assumptions.}

\section{Bayesian estimation of W}

In this paper we use a Bayesian estimation approach to obtain estimates and inference on the unknown quantities $\rho$, $\boldsymbol{\beta}$, $\sigma^2$, as well as the elements of $\boldsymbol{\Omega}$. After eliciting suitable priors for the unknown parameters,  we employ a computationally efficient MCMC algorithm. 

Let $p(\omega_{ij}=1)$  denote the prior belief in including the $ij$th element of the spatial weight matrix. Conversely, for a proper prior specification the prior probability of exclusion is then simply given by $p(\omega_{ij}=0)=1-p(\omega_{ij}=1)$. With $\boldsymbol{\Omega}_{-ij}$ denoting the elements of the neighbourhood matrix without $\omega_{ij}$, the posterior probabilities of $\omega_{ij}=1$  and $\omega_{ij}=0$ conditional on all other parameters are given by:
\begin{alignat}{1}
\begin{aligned}
\label{eq:p1p0}
p(\omega_{ij}=1| \boldsymbol{\Omega}_{-ij},\boldsymbol{\beta},\sigma^2,\rho,\mathcal{D}) \propto  p(\omega_{ij}=1)|\boldsymbol{S}_1|\exp\left[-\frac{1}{2\sigma^2} (\boldsymbol{S}_{1}\boldsymbol{Y}-\boldsymbol{X\beta})'(\boldsymbol{S}_{1}\boldsymbol{Y}-\boldsymbol{X\beta})\right]\phantom{,}\\
p(\omega_{ij}=0|\boldsymbol{\Omega}_{-ij},\boldsymbol{\beta},\sigma^2,\rho,\mathcal{D}) \propto  p(\omega_{ij}=0)|\boldsymbol{S}_0|\exp\left[-\frac{1}{2\sigma^2} (\boldsymbol{S}_{0}\boldsymbol{Y}-\boldsymbol{X\beta})'(\boldsymbol{S}_{0}\boldsymbol{Y}-\boldsymbol{X\beta})\right],
\end{aligned}
\end{alignat}
where $\boldsymbol{S}_{1}$ and $\boldsymbol{S}_{0}$ are given by $\boldsymbol{S}$ through updating  the spatial weight matrix $\boldsymbol{W}$ via setting $\omega_{ij}=1$ and $\omega_{ij}=0$, respectively.\footnote{To reduce the dimensionality of the parameter space, an interesting alternative might be the assumption of a symmetric $\boldsymbol{\Omega}$, which halves the number of free elements in the spatial weight matrix. This assumption can be imposed in the way by simply simultaneously updating $\omega_{ij}=\omega_{ji}$, respectively.} Using the law of total probability, it is straightforward to show that the resulting conditional posterior for $\omega_{ij}$ is Bernoulli:
\begin{eqnarray}
\label{eq:condpostw}
p(\omega_{ij}| \boldsymbol{\Omega}_{-ij},\boldsymbol{\beta},\sigma^2,\rho,\mathcal{D})\sim\mathcal{BER}\left(\frac{\bar{p}^{(1)}_{ij}}{\bar{p}^{(0)}_{ij}+\bar{p}^{(1)}_{ij}}\right),
\end{eqnarray}
with $\bar{p}^{(1)}_{ij}=p(\omega_{ij}=1| \boldsymbol{\Omega}_{-ij},\boldsymbol{\beta},\sigma^2,\rho,\mathcal{D})$ and $\bar{p}^{(0)}_{ij}=p(\omega_{ij}=0| \boldsymbol{\Omega}_{-ij},\boldsymbol{\beta},\sigma^2,\rho,\mathcal{D})$ given in Eq. (\ref{eq:p1p0}). Since the conditional posterior follows a convenient and well-known form, efficient Gibbs sampling can be employed.

A Bayesian estimation framework requires elicitation of  a prior on $\boldsymbol{\Omega}$. Obvious candidates  are independent Bernoulli priors on the unknown indicators $\omega_{ij}$:
\begin{equation}
\label{eq:fixedprior}
p(\omega_{ij})\sim\mathcal{BER}\left(\underline{p}_{ij}\right),
\end{equation}
where $\underline{p}_{ij}$ denotes the prior inclusion probability of $\omega_{ij}$, $p(\omega_{ij}=1)=\underline{p}_{ij}$. Conversely, the prior probability of exclusion then simply takes the form $p(\omega_{ij}=0)=1-\underline{p}_{ij}$. 

A natural prior choice would involve setting $\underline{p}_{ij}=\underline{p}=1/2$ for $i\neq j$, and zero otherwise, which  implies that each off-diagonal element in $\boldsymbol{\Omega}$  has an equal prior chance of being included. However, in many cases a researcher has possible a priori information on the underlying structure of the spatial weight matrix. The following stylized examples demonstrate how to incorporate such information in a flexible and straightforward way.


\begin{figure}[!ht]
\caption{Some stylized prior examples for $\boldsymbol{W}$ in a linear city}
\label{fig:priors}
\centering
\begin{minipage}[b]{.45\linewidth}
\subcaption*{(A) Exogenous $\boldsymbol{W}$}\vspace{-0.75cm}\label{fig:priora}
\centering \includegraphics[scale=0.35]{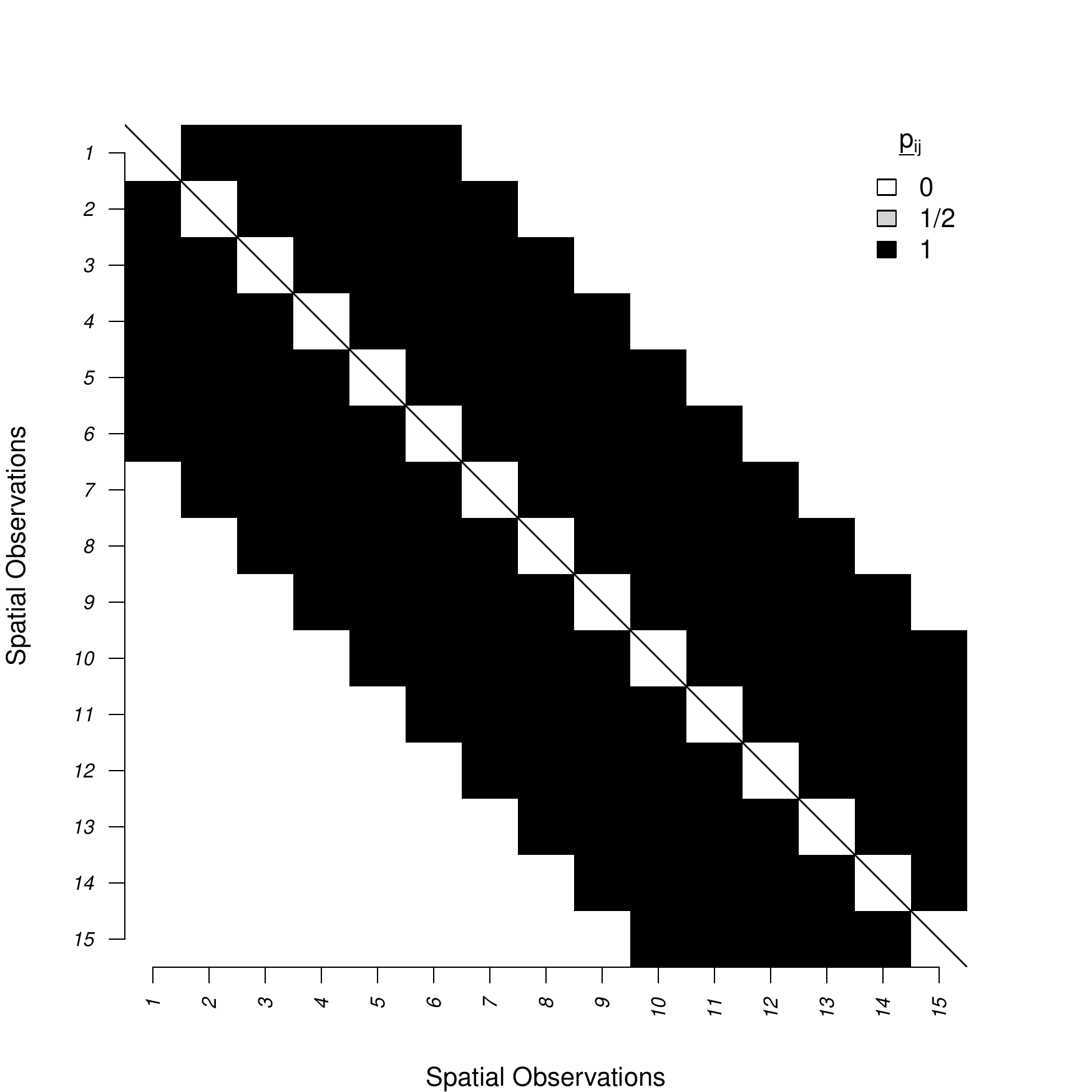}
\end{minipage}%
\begin{minipage}[b]{.45\linewidth}
\subcaption*{(B) Fixed ($\underline{p}=1/2$)}\vspace{-0.75cm}\label{fig:priorb}
\centering \includegraphics[scale=0.35]{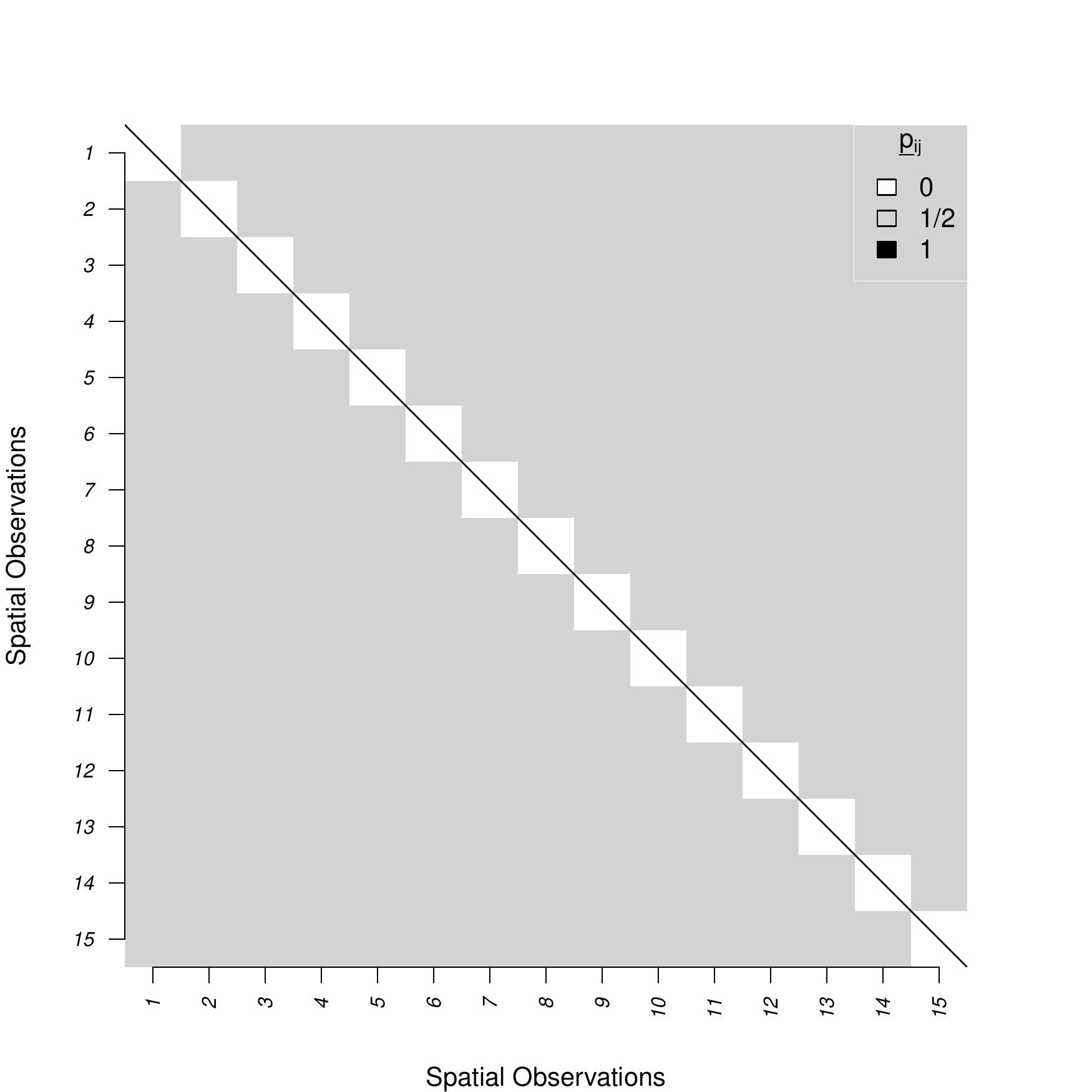}
\end{minipage}\\\vspace{0.5cm}
\begin{minipage}[b]{.45\linewidth}
\subcaption*{(C) Spatial prior}\vspace{-0.75cm}\label{fig:priorc}
\centering \includegraphics[scale=0.35]{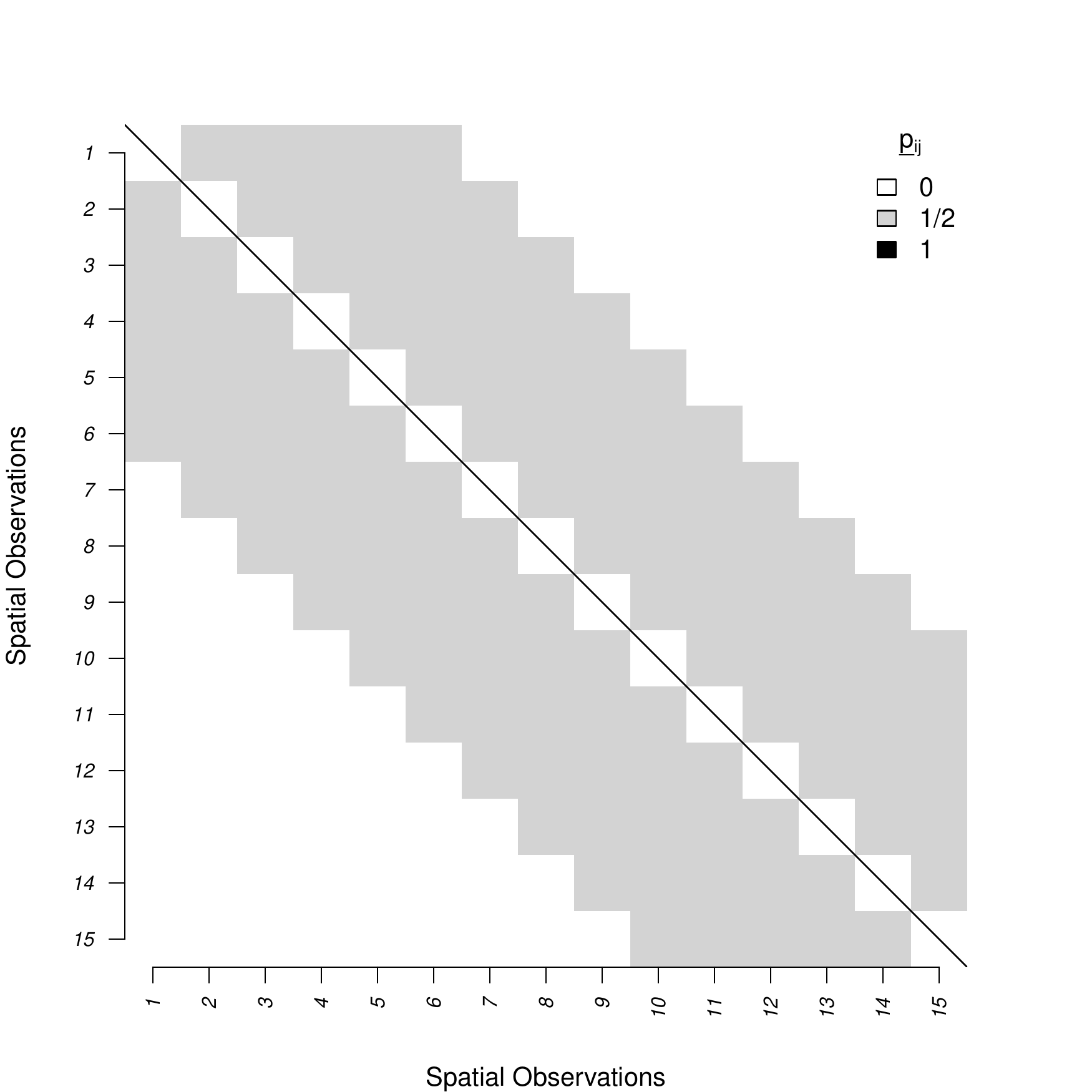}
\end{minipage}
\begin{minipage}[b]{.45\linewidth}
\subcaption*{(D) Spatial prior: combining two $\boldsymbol{W}$'s}\vspace{-0.75cm}\label{fig:priord}
\centering \includegraphics[scale=0.35]{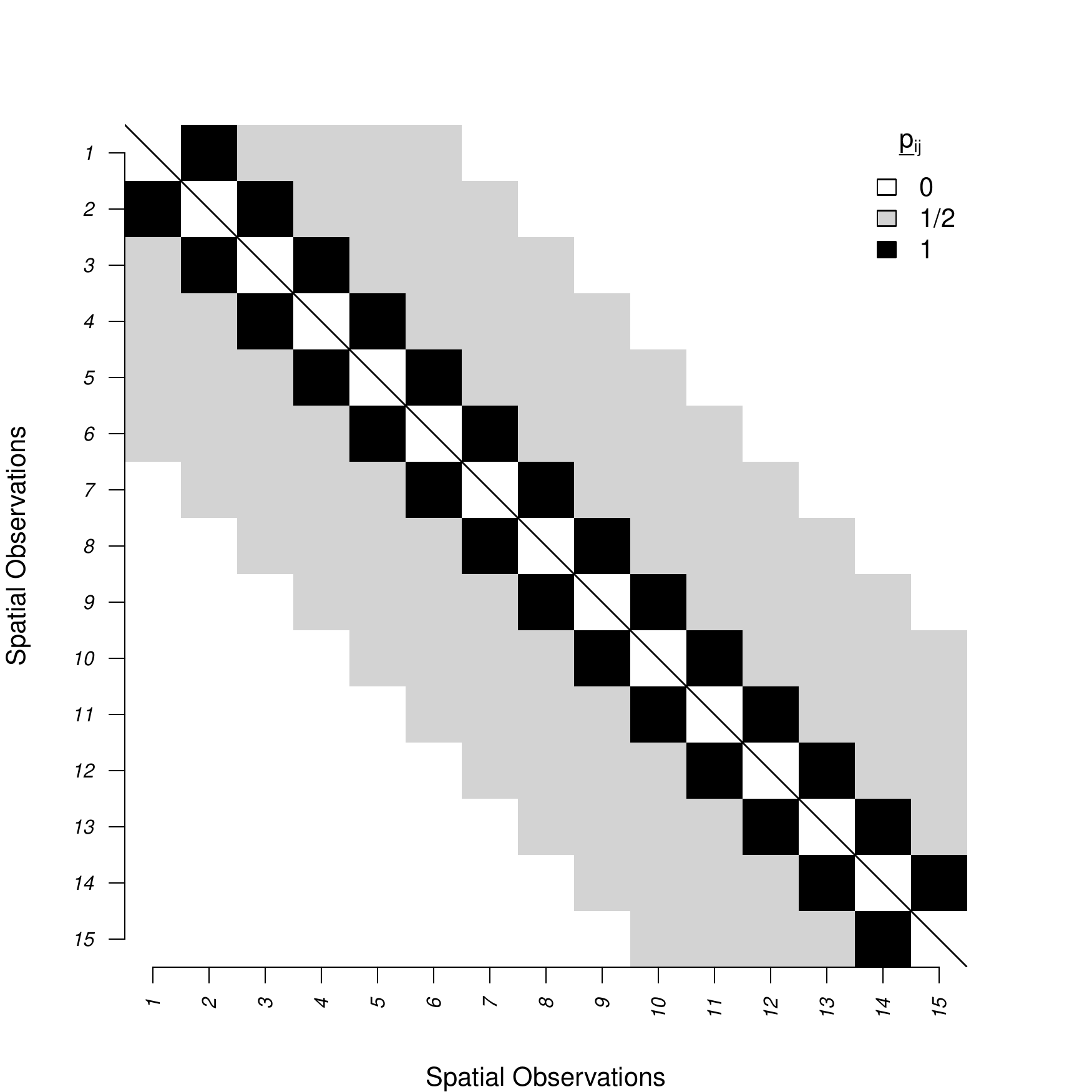}
\end{minipage}\\
\begin{minipage}{12.5cm}~\\
\scriptsize \textbf{Notes}: Alternative prior setups for a linear city of $N=15$ spatial observations. Case (A) shows a prior specification without any prior uncertainty on the spatial links. This setup implies an exogenous  $\boldsymbol{W}$ and no estimation of the weights is involved. Case (B) involves no spatial prior information and each element has a prior probability of inclusion $\underline{p}_{ij}=1/2 \, \forall i\neq j$. Case (C) shows uncertainty of the linkages in $\boldsymbol{W}$ only within a certain spatial domain. Case (D) is a stylized prior specification considering uncertainty among two (or more) weight matrices, with setting $p_{ij}=1$ in regions where the two matrices overlap.  
\end{minipage}%
\end{figure}

Figure \ref{fig:priors} illustrates the flexibility of prior elicitation for $\boldsymbol{\Omega}$ in the case of a "linear city" with $N=15$ equidistant regions. Case (A) in the figure shows a prior specification without any prior uncertainty on the elements of $\boldsymbol{W}$ by setting $\underline{p}_{ij}=1$ if $i$ and $j$ are considered as neighbours and zero otherwise. In this case, no estimation on the spatial links is involved and the model reduces to a standard SAR model with an exogenously given $\boldsymbol{W}$ (in this example, a distance band specification).

Case (B) depicts the opposite case where no prior spatial information is available. Specifically, this case considers full estimation of all $N^2-N$ potential links with respective prior inclusion probability $\underline{p}_{ij}=1/2$ for $i\neq j$.

Subplots (C) and (D) in Figure \ref{fig:priors}  depict prior setups where a priori spatial information is available to the researcher, but associated with uncertainty. Case (C) illustrates a prior where the general spatial domain is assumed as being a priori known, but uncertainty over specific linkages exists. In empirical practice, spatial weight matrices based on geographic information are often viewed as being preferable due to exogeneity assumptions to (socio-)economic data. The illustrated prior specification follows this idea by still allowing for uncertainty and flexibility among the spatial neighbourhood. 

Recent contributions to spatial econometric literature propose selecting (\citealt{Piribauer2015}) or combining (\citealt{debarsy2018flexible}) multiple exogenous  spatial weight matrices. Case (D) follows a similar idea by depicting a mixture of a distance band and a contiguity matrix (i.e. neighbourhood if regions share a common border). The intersecting elements of the two spatial structures (resulting in a contiguity matrix) are assumed as being included by setting $p_{ij}=1$.

\subsection*{Hierarchical prior setups and sparsity}

The prior structure in Eq. (\ref{eq:fixedprior}) involves \emph{fixed} inclusion probabilities $\underline{p}$, which implies that the number of neighbours of observation $i$ follows a Binomial distribution $\sum_{l=1}^{N-1}\omega_{il}\sim\mathcal{BN}(N-1,\underline{p})$ with a prior expected number of neighbours of $(N-1)\underline{p}$. However, such a prior structure has the potential undesirable effect of promoting a relatively large number of neighbours. For example, when $\underline{p}=1/2$, the prior expected number of neighbours is $(N-1)/2$, since combinations of $\omega_{ij}$ resulting in such a neighbourhood size are dominant in number. 

To put more prior weight on parsimonious neighbourhood structures and therefore promote sparsity in the adjacency matrix, one may explicitly account for the number of linkages in each row of the adjacency matrix $\boldsymbol{\omega}_{i}=\left[\omega_{i1},\dots,\omega_{iN}\right]'$. We consider a flexible prior structure on the number of neighbours $\sum \boldsymbol{\omega}_i$ that corresponds to a beta-binomial distribution $\mathcal{BB}(N-1,\underline{a}_{\omega},\underline{b}_{\omega})$ with  two prior hyperparameters $\underline{a}_{\omega},\underline{b}_{\omega}>0$. The beta-binomial distribution is the result of treating the prior inclusion probability $\underline{p}$ as \emph{random} (rather than being fixed) by placing a hierarchical beta prior on it. For $\omega_{ij}$, the resulting prior can be written as follows:


\begin{equation}
\label{eq:hierarchicalprior}
p(\omega_{ij})\propto \Gamma\left(\underline{a}_w+\sum\boldsymbol{\omega}_i\right)\Gamma\left(\underline{b}_{\omega}+(N-1)-\sum\boldsymbol{\omega}_i\right),
\end{equation}
where $\Gamma(\cdot )$ denotes the Gamma function, and $\underline{a}_{\omega}$ and $\underline{b}_{\omega}$  are prior hyperparameters.

In the case of $\underline{a}_{\omega}=\underline{b}_{\omega}=1$, the prior takes the form of a discrete uniform distribution over the number of neighbours. By fixing $\underline{a}_{\omega}=1$, we follow \cite{leysteel2009} and anchor the prior expected number of neighbours $\underline{m}$ via $\underline{b}_{\omega}=[(N-1)-\underline{m}]/\underline{m}$. 


\section{Bayesian MCMC estimation of the model} 
This section presents the Bayesian MCMC  estimation algorithm for the proposed modelling framework. Estimation is carried out using an efficient Gibbs sampling scheme. The only exception is the sampling step for the spatial (scalar) autoregressive parameter $\rho$, where we propose using a standard griddy Gibbs step.\footnote{A random walk Metropolis-Hastings step for $\rho$ might be employed as an alternative.} The sampling scheme involves the following steps:
\begin{enumerate}
\item[I.] Set starting values for the parameters (e.g. by sampling from the prior distributions)
\item[II.] Sequentially update the parameters by subsequently sampling from the conditional posterior distributions presented in this section.
\end{enumerate}
Step II. is repeated for $B$ times after discarding the first $B_0$ draws as burn-ins.

\subsection*{Sampling $\boldsymbol{\beta}$ and $\sigma^2$}
For the slope parameters $\boldsymbol{\beta}$ and the error variance $\sigma^2$ we use common Normal and inverted Gamma prior specifications, respectively. Specifically,  $p(\boldsymbol{\beta})\sim\mathcal{N}(\boldsymbol{0},\underline{\boldsymbol{V}}_\beta)$ and $p(\sigma^2)\sim\mathcal{IG}(\underline{a}_{\sigma^2},\underline{b}_{\sigma^2}$), where  $\underline{\boldsymbol{V}}_\beta$, $\underline{a}_{\sigma^2}$, and $\underline{b}_{\sigma^2}$ denote prior hyperparameters. 


The resulting conditional posterior distribution is Gaussian and of well-known form (see, for example, \citealt{LeSage2009}):
\begin{eqnarray}
p(\boldsymbol{\beta}|\sigma^2,\rho,\boldsymbol{\Omega},\mathcal{D}) &\sim& \mathcal{N}(\bar{\boldsymbol{b}}_\beta,\bar{\boldsymbol{V}}_\beta)\\
\bar{\boldsymbol{b}}_\beta&=&\sigma^{-2}\bar{\boldsymbol{V}}_\beta\boldsymbol{X}'\boldsymbol{SY} \notag \\
\bar{\boldsymbol{V}}_\beta&=&\left(\sigma^{-2}\boldsymbol{X}'\boldsymbol{X}+\underline{\boldsymbol{V}}_\beta^{-1}\right)^{-1}. \notag
\end{eqnarray}
The conditional posterior of $\sigma^2$ is inverted Gamma:
\begin{eqnarray}
p(\sigma^2 | \boldsymbol{\beta},\rho,\boldsymbol{\Omega},\mathcal{D}) &\sim& \mathcal{IG}(\bar{a}_{\sigma^2},\bar{b}_{\sigma^2})\\
\bar{a}_{\sigma^2}&=&\underline{a}_{\sigma^2}+NT/2 \notag\\
\bar{b}_{\sigma^2}&=&\underline{b}_{\sigma^2}+(\boldsymbol{SY}-\boldsymbol{X\beta})'(\boldsymbol{SY}-\boldsymbol{X\beta}). \notag
\end{eqnarray}

\subsection*{Sampling $\rho$} 
For the spatial parameter $\rho$, we use a standard Beta distribution \citep[see][p. 142]{LeSage2009}. The conditional posterior is given by:
\begin{equation}
p(\rho | \boldsymbol{\beta},\sigma^2,\boldsymbol{\Omega},\mathcal{D}) \propto p(\rho) |\boldsymbol{S}| \exp\left[-\frac{1}{2\sigma^2} (\boldsymbol{SY}-\boldsymbol{X\beta})'(\boldsymbol{SY}-\boldsymbol{X\beta})\right]. \label{eq:update_rho}
\end{equation}
Note that the conditional posterior for $\rho$ does not follow a well-known form and thus requires alternative sampling techniques. We follow \cite{LeSage2009} and use a griddy-Gibbs step (\citealt{Ritter1992}) to sample $\rho$.\footnote{Since the support for $\rho$ is limited, the griddy-Gibbs approach (or sometimes inversion approach) relies on univariate numerical integration techniques of the conditional posterior for $\rho$ and uses the cumulative density function for producing draws of $\rho$. A Metropolis-Hastings step may be used as a standard alternative, but these typically produce less efficient draws with poorer mixing properties (see also \citealt{LeSage2009}).}



\subsection*{Sampling the elements of the adjacency matrix $\boldsymbol{\Omega}$}
As discussed in the previous section, we propose two alternative prior specifications for the unknown indicators of the spatial weight matrix $\omega_{ij}.$ First, an independent Bernoulli prior structure with fixed inclusion probabilities (\ref{eq:fixedprior}). Second, a hierarchical prior structure which treats the inclusion probabilities as random (\ref{eq:hierarchicalprior}). After eliciting the prior, the binary indicators $\omega_{ij}$ can be sequentially sampled in random order from a Bernoulli distribution with conditional posterior given in (\ref{eq:condpostw}). 

\subsection*{Fast computation of the determinant terms \label{sec:computational}}

For the Bayesian MCMC algorithm, it is worth noting that repeated sampling from  Eq. (\ref{eq:condpostw}) is required. However, this requires evaluating the conditional probabilities $p(\omega_{ij} = 1|\cdot)$ and $p(\omega_{ij} = 0|\cdot)$ in Eq. (\ref{eq:p1p0}). The main computational difficulty lies in the calculation of the determinants $|\boldsymbol{S}_0|$ and $|\boldsymbol{S}_1|$, which has to be carried out per Gibbs sampling step for the $N^2-N$ unknown elements of the spatial adjacency matrix. The computational costs associated with direct calculation of these determinants steeply rises with $N$ -- in fact by a factor of $\mathcal{O}(N^3)$. This makes direct evaluation of the determinant prohibitively expensive, especially for large values of $N$. To avoid direct evaluation, we provide computationally efficient updates for the determinant, allowing for estimation of models with larger sample sizes. 

It is worth noting that it is not necessary to directly calculate the determinant of the $NT \times NT$ matrix $\boldsymbol{S}_z$ (with $z \in \{0,1\}$). Only the determinant of the $N \times N$ matrix $\boldsymbol{A}_z = \boldsymbol{I}_N - \rho \boldsymbol{W}_z$ needs to be updated, since $|\boldsymbol{S}_z|  = |\boldsymbol{I}_T \otimes \boldsymbol{A}_z| = |\boldsymbol{A}_z|^T$.  Here, $\boldsymbol{W}_z$ denotes the spatial weight matrix obtained by setting $\omega_{ij}=1$ and $\omega_{ij} = 0$, respectively.

Direct evaluation of $|\boldsymbol{A}_z|$ can be largely avoided, since updating $\omega_{ij}$ changes only the $i$-th row of $\boldsymbol{A}$, if we do not restrict $\boldsymbol{\Omega}$ to be symmetric (we will address this case shortly). To illustrate, let $\boldsymbol{\Omega}^{(c)}$ denote the current -- to be updated -- spatial adjacency matrix, and $\boldsymbol{W}^{(c)}$ the associated spatial weight matrix with determinant $|\boldsymbol{A}^{(c)}| = |\boldsymbol{I}_N - \rho \boldsymbol{W}^{(c)}|$. Using the so-called matrix determinant lemma, we can efficiently calculate:
\begin{align}
|\boldsymbol{A}_z| = |\boldsymbol{A}^{(c)} + \boldsymbol{\nu}_i \boldsymbol{\delta}_i'| = \left\lbrace1 + \boldsymbol{\delta}_i' (\boldsymbol{A}^{(c)})^{-1} \boldsymbol{\nu}_i \right\rbrace| \boldsymbol{A}^{(c)}|. \label{eq:mat_det_lemma}
\end{align}
$\boldsymbol{\nu}_i$ is an $N \times 1$ vector of zeros, except for its $i$-th entry, which is unity. The $N\times 1$ vector $\boldsymbol{\delta}_i$ contains the differences between the $i$-th row of $\boldsymbol{A}_z$ and the $i$-th row of $\boldsymbol{A}^{(c)}$. 

It becomes clear that Eq. (\ref{eq:mat_det_lemma}) provides a computationally cheap way for updating the determinant $|\boldsymbol{A}_z|$, conditional on $|\boldsymbol{A}^{(c)}|$ and $\left(\boldsymbol{A}^{(c)}\right)^{-1}$. This implies that during the MCMC procedure, for each update of $\omega_{ij}$, we have to keep track of the determinant (for which Eq. (\ref{eq:mat_det_lemma}) provides a simple update) and the inverse of $\boldsymbol{A}_z$. Direct evaluation of $\boldsymbol{A}_z^{-1}$ is -- similar to direct evaluation of the determinant -- prohibitively expensive for moderate to large $N$, since it has to be carried out for each unknown element of $\boldsymbol{\Omega}$. However, we can rely on the so-called Sherman-Morrison formula to avoid direct evaluation of the matrix inverse: 
\begin{align}
\boldsymbol{A}_z^{-1} = \left( \boldsymbol{A}^{(c)} + \boldsymbol{\nu}_i \boldsymbol{\delta}_i' \right)^{-1} = \left(\boldsymbol{A}^{(c)}\right)^{-1} - \frac{\left(\boldsymbol{A}^{(c)}\right)^{-1} \boldsymbol{\nu}_i \boldsymbol{\delta}_i' \left(\boldsymbol{A}^{(c)}\right)^{-1}}{1 + \boldsymbol{\delta}_i' \left(\boldsymbol{A}^{(c)}\right)^{-1} \boldsymbol{\nu}_i} . \label{eq:sherman_morrison}
\end{align}
Combining the formulas in Eqs. (\ref{eq:mat_det_lemma}) and (\ref{eq:sherman_morrison}) thus provides a numerically cheap and viable way to update the elements of the  spatial adjacency matrix.\footnote{Note the implication that an update of $\rho$ necessitates a direct evaluation of the determinant $|\boldsymbol{A}|$ and the matrix inverse $\boldsymbol{A}^{-1}$, as in this case no convenient equations exist. An update of $\rho$, however, has to be performed only once per Gibbs step, as opposed to the $N^2-N$ updates necessary for $\boldsymbol{\Omega}$, thus justifying the relatively higher computational costs.} 

The binary nature of $\omega_{ij}$ can be exploited for additional computational gains. Either $\boldsymbol{A}_0$ or $\boldsymbol{A}_1$ always exactly equals $\boldsymbol{A}^{(c)}$ and thus its determinant and inverse is already known. This only necessitates calculating $|\boldsymbol{A}_z|$ and $(\boldsymbol{A}_z)^{-1}$  for only $z=1$ or for $z=0$, but not both.

If a symmetric spatial adjacency matrix $\boldsymbol{\Omega}$ is assumed, the update process remains generally the same, however the determinant and matrix inverse updates have to be performed iteratively. In this case, both $\omega_{ij}$ and $\omega_{ji}$ (for $i \neq j$) are set to either $1$ or $0$. Thus, both the $i$-th and the $j$-th row of $\boldsymbol{A}_z$ differ from $\boldsymbol{A}^{(c)}$. Following the notation in the non-symmetric case, let us denote the differences between these rows as $\boldsymbol{\delta}_i$ and $\boldsymbol{\delta}_j$. To obtain an update of $|\boldsymbol{A}_z|$ and $\boldsymbol{A}_z^{-1}$, we first evaluate Eqs. (\ref{eq:mat_det_lemma}) and (\ref{eq:sherman_morrison}), based on $\boldsymbol{\delta}_i$, $\boldsymbol{\nu}_i$, $|\boldsymbol{A}^{(c)}|$, and $(\boldsymbol{A}^{(c)})^{-1}$. Using  the resulting determinant and matrix inverse, as well as $\boldsymbol{\nu}_j$, and $\boldsymbol{\delta}_j$, we again evaluate Eqs. (\ref{eq:mat_det_lemma}) and (\ref{eq:sherman_morrison}), which yield $|\boldsymbol{A}_z|$ and $\boldsymbol{A}_z^{-1}$.

\section{Simulation study}

To assess the accuracy of our proposed approach, we evaluate its performance in a  Monte Carlo study. Our benchmark data generating process comprises two randomly generated explanatory variables, as well as  spatial unit and time fixed effects:
\begin{align*}
\tilde{\boldsymbol{y}}_t= \tilde{\rho}\widetilde{\boldsymbol{W}}\tilde{\boldsymbol{y}}_t+\tilde{\boldsymbol{\mu}}+\tilde{\tau}_t+\tilde{\boldsymbol{Z}}_t\tilde{\boldsymbol{\beta}}_0+\tilde{\boldsymbol{\varepsilon}}_t.
\end{align*}
To maintain succinct notation, we denote the simulated values in the Monte Carlo study with a tilde. The matrix of explanatory variables $\tilde{\boldsymbol{Z}}_t$ is defined as $\tilde{\boldsymbol{Z}}_t = [\tilde{{z}}_{1t},\tilde{{z}}_{2t}]$, where both $\tilde{{z}}_{1t}$ and $\tilde{{z}}_{2t}$ are normally distributed with zero mean and variance of one, $q_0=2$. The corresponding vector of coefficients is defined as  $\tilde{\boldsymbol{\beta}}_0 = [-1,1]'$. The vector of residuals $\tilde{\boldsymbol{\varepsilon}}_t$ is generated from a normal distribution with zero mean and $\tilde{\sigma}^2 = 0.5$. The fixed effects parameters $\tilde{\boldsymbol{\mu}}$ and $\tilde{\tau}_t$ are randomly generated from a standard normal distribution.

The row-stochastic spatial weight matrix $\widetilde{\boldsymbol{W}}$ is based on an adjacency matrix $\widetilde{\boldsymbol{\Omega}}$, which is generated from an $N/20$  nearest neighbour specification, by additionally assuming symmetry of the weight matrix prior to row-standardization.\footnote{More specifically, $\widetilde{\boldsymbol{\Omega}} = ( \widetilde{\boldsymbol{\Omega}}_0' + \widetilde{\boldsymbol{\Omega}}_0 )/2$ where $\widetilde{\boldsymbol{\Omega}}_0$ is a $N/20$ nearest neighbour adjacency matrix.} The nearest neighbour specification is based on a randomly generated spatial location pattern, sampled from a normal distribution with zero mean and unity variance. In the Monte Carlo study we vary $T \in \{10,40\}$ and $N \in \{20,100\}$. Additionally, we vary the strength of spatial dependence $\tilde{\rho} \in \{0.3,0.5,0.8\}$.


For the Monte Carlo simulation study, we compare the following prior setups:
\begin{enumerate}
\item  \textit{Fixed} ($\underline{p} = 1/2$) prior: this prior corresponds to the fixed Bernoulli prior specification in Eq. (\ref{eq:fixedprior}), where we set $\underline{p}=1/2$.
\item \textit{Sparsity} ($\underline{m} = (N-1)/2$) prior: this is analogous to the prior setup in Eq. (\ref{eq:hierarchicalprior}), with $\underline{a}_{\omega} = \underline{b}_{\omega} = 1$. This prior setup corresponds to a discrete uniform distribution over the number of neighbours.
\item \textit{Sparsity} ($\underline{m} = N/10$) prior: this prior setup corresponds to Eq. (\ref{eq:hierarchicalprior}), with $\underline{a}_{\omega} = 1$ and $\underline{b}_{\omega} = [(N − 1) − \underline{m}]/\underline{m}$. We set the number of a priori expected neighbours to $\underline{m} = N/10$. This prior setup thus imposes more sparsity in $\boldsymbol{\Omega}$ as compared to the former.
\end{enumerate}
For all prior specifications under scrutiny, we consider two alternative estimation setups by assuming that the adjacency matrix is either \textit{symmetric} or \textit{non-symmetric}.\footnote{However, a direct comparison of the results between symmetric and non-symmetric specifications does not appear reasonable, since the adjacency matrix in the data generating process is assumed symmetric.} We moreover report the predictive performance of two alternative specifications using exogenous weight matrices. In these cases the employed weights are based on the true (symmetric) adjacency matrix by fixing the accuracy to the 99\% and 95\% level, respectively. We simulate such cases by randomly switching 1\% and 5\% of the elements in the true binary adjacency matrix $\widetilde{\boldsymbol{\Omega}}$, respectively. The resulting exogenous adjacency matrices thus result in exactly 99\% and 95\% overlap in the binary observations with the true adjacency matrix, while maintaining the same level of sparsity. 

The prior setup for our remaining parameters is as follows. We assume a Gaussian prior for $\boldsymbol{\beta}$ with zero mean and a variance of $100$. We use an inverse gamma prior for $\sigma^2$ with rate and shape parameters $0.01$. The prior for the spatial autoregressive parameter $\rho$ is a symmetric Beta specification with shape and rate parameters equal to $1.01$. The chosen priors can thus be considered highly non-informative.

In Table \ref{tab:MC1} we use several criteria to evaluate the performance of the alternative specifications. For the  spatial autoregressive and the slope parameters we report the well-known root mean squared error (RMSE). For assessing the ability to estimating the spatial adjacency matrix, we use the measure of accuracy.  The accuracy measure is defined as the sum of correctly identified unknown elements, divided by the number of total elements to be estimated. This measure is calculated separately for each posterior draw. The reported value is an average over all posterior draws and Monte Carlo iterations.

\begin{table}[!h]
\begin{threeparttable}[tbp]
  \centering
  \caption{Monte Carlo simulation results}
  \scriptsize
 \renewcommand{\arraystretch}{0.8}
\begin{tabularx}{\textwidth}{cccc YYYc@{\hskip 0.2cm} YYYc@{\hskip 0.2cm} cc}
\toprule
\multirow{3}[4]{*}{} & \multirow{3}[4]{*}{$N$} & \multirow{3}[4]{*}{$T$} & \multirow{3}[4]{*}{$\tilde{\rho}$} & \multicolumn{3}{c}{Non-symmetric} &    & \multicolumn{3}{c}{Symmetric} &    & \multicolumn{2}{c}{Exogenous} \\
\cmidrule{5-11}   &    &    &    & \multicolumn{1}{c}{Fixed} & \multicolumn{1}{c}{Sparsity} & \multicolumn{1}{c}{Sparsity} &    & \multicolumn{1}{c}{Fixed} & \multicolumn{1}{c}{Sparsity} & \multicolumn{1}{c}{Sparsity} &    & \multicolumn{2}{c}{$\boldsymbol{W}$} \\
   &    &    &    & \multicolumn{1}{c}{$\underline{p} = 1/2$} & \multicolumn{1}{c}{$\underline{m}=N/2$} & \multicolumn{1}{c}{$\underline{m} = N/10$} &    & \multicolumn{1}{c}{$\underline{p} = 1/2$} & \multicolumn{1}{c}{$\underline{m}=n/2$} & \multicolumn{1}{c}{$\underline{m} = n/10$} &    & \multicolumn{1}{c}{$0.99$} & \multicolumn{1}{c}{$0.95$} \\
\midrule
\multirow{12}[4]{*}{\begin{sideways}RMSE($\boldsymbol{\beta}$)\end{sideways}} & \multirow{6}[2]{*}{20} & \multirow{3}[1]{*}{40} & 0.3 & 0.193 & \textbf{0.161} & 0.163 &    & \textbf{0.162} & 0.163 & 0.164 &    & 0.168 & 0.176 \\
   &    &    & 0.5 & 0.172 & 0.173 & \textbf{0.172} &    & 0.170 & \textbf{0.169} & 0.171 &    & 0.179 & 0.216 \\
   &    &    & 0.8\bigstrut[b] & \textbf{0.169} & 0.169 & 0.169 &    & \textbf{0.165} & 0.166 & 0.167 &    & 0.287 & 0.553 \\
   &    & \multirow{3}[1]{*}{10} & 0.3 & 0.234 & 0.207 & \textbf{0.198} &    & 0.203 & 0.182 & \textbf{0.181} &    & 0.192 & 0.204 \\
   &    &    & 0.5 & 0.257 & 0.210 & \textbf{0.206} &    & \textbf{0.189} & 0.191 & 0.190 &    & 0.206 & 0.253 \\
   &    &    & 0.8\bigstrut[b] & 0.217 & \textbf{0.216} & 0.217 &    & 0.205 & \textbf{0.204} & 0.206 &    & 0.371 & 0.658 \\
\cmidrule{2-14}   & \multirow{6}[2]{*}{100} & \multirow{3}[1]{*}{40} & 0.3 & 0.098 & 0.099 & \textbf{0.097} &    & 0.099 & 0.099 & \textbf{0.098} &    & 0.079 & 0.080 \\
   &    &    & 0.5 & 0.144 & 0.088 & \textbf{0.083} &    & 0.145 & 0.114 & \textbf{0.076} &    & 0.084 & 0.086 \\
   &    &    & 0.8\bigstrut[b] & 0.154 & \textbf{0.087} & 0.088 &    & \textbf{0.073} & 0.081 & 0.081 &    & 0.089 & 0.141 \\
   &    & \multirow{3}[1]{*}{10} & 0.3 & \textbf{0.111} & 0.112 & 0.111 &    & \textbf{0.111} & 0.111 & 0.112 &    & 0.092 & 0.093 \\
   &    &    & 0.5 & 0.135 & 0.118 & \textbf{0.104} &    & 0.135 & 0.136 & \textbf{0.118} &    & 0.088 & 0.094 \\
   &    &    & 0.8\bigstrut[b] & 0.346 & 0.143 & \textbf{0.140} &    & 0.254 & 0.102 & \textbf{0.102} &    & 0.100 & 0.151 \\
\midrule
\multirow{12}[4]{*}{\begin{sideways}RMSE($\rho$)\end{sideways}} & \multirow{6}[2]{*}{20} & \multirow{3}[1]{*}{40} & 0.3 & 0.199 & \textbf{0.029} & 0.031 &    & 0.030 & \textbf{0.029} & 0.030 &    & 0.034 & 0.060 \\
   &    &    & 0.5 & \textbf{0.035} & 0.040 & 0.042 &    & \textbf{0.035} & 0.035 & 0.035 &    & 0.039 & 0.083 \\
   &    &    & 0.8\bigstrut[b] & \textbf{0.021} & 0.021 & 0.022 &    & 0.018 & \textbf{0.018} & 0.018 &    & 0.084 & 0.177 \\
   &    & \multirow{3}[1]{*}{10} & 0.3 & 0.237 & 0.152 & \textbf{0.094} &    & 0.291 & 0.147 & \textbf{0.106} &    & 0.058 & 0.080 \\
   &    &    & 0.5 & 0.155 & 0.060 & \textbf{0.054} &    & 0.109 & 0.053 & \textbf{0.051} &    & 0.059 & 0.114 \\
   &    &    & 0.8\bigstrut[b] & \textbf{0.027} & 0.032 & 0.032 &    & \textbf{0.028} & 0.028 & 0.029 &    & 0.097 & 0.179 \\
\cmidrule{2-14}   & \multirow{6}[2]{*}{100} & \multirow{3}[1]{*}{40} & 0.3 & 0.280 & 0.283 & \textbf{0.277} &    & \textbf{0.279} & 0.283 & 0.287 &    & 0.027 & 0.033 \\
   &    &    & 0.5 & 0.447 & 0.109 & \textbf{0.101} &    & 0.446 & 0.353 & \textbf{0.220} &    & 0.021 & 0.054 \\
   &    &    & 0.8\bigstrut[b] & 0.148 & \textbf{0.044} & 0.047 &    & 0.049 & \textbf{0.024} & 0.024 &    & 0.034 & 0.097 \\
   &    & \multirow{3}[1]{*}{10} & 0.3 & \textbf{0.242} & 0.256 & 0.268 &    & \textbf{0.245} & 0.252 & 0.274 &    & 0.050 & 0.062 \\
   &    &    & 0.5 & 0.373 & 0.176 & \textbf{0.141} &    & \textbf{0.371} & 0.391 & 0.404 &    & 0.041 & 0.074 \\
   &    &    & 0.8\bigstrut[b] & 0.473 & \textbf{0.106} & 0.110 &    & 0.169 & 0.141 & \textbf{0.137} &    & 0.044 & 0.105 \\
\midrule
\multirow{12}[4]{*}{\begin{sideways}Accuracy $\boldsymbol{\Omega}$\end{sideways}} & \multirow{6}[2]{*}{20} & \multirow{3}[1]{*}{40} & 0.3 & 0.648 & 0.930 & \textbf{0.954} &    & 0.963 & 0.982 & \textbf{0.983} &    & 0.990 & 0.950 \\
   &    &    & 0.5 & 0.983 & 0.988 & \textbf{0.989} &    & 0.998 & 0.998 & \textbf{0.998} &    & 0.990 & 0.950 \\
   &    &    & 0.8\bigstrut[b] & \textbf{0.995} & 0.995 & 0.995 &    & 0.999 & \textbf{1.000} & 0.999 &    & 0.990 & 0.950 \\
   &    & \multirow{3}[1]{*}{10} & 0.3 & 0.554 & 0.752 & \textbf{0.866} &    & 0.679 & 0.875 & \textbf{0.904} &    & 0.990 & 0.950 \\
   &    &    & 0.5 & 0.734 & 0.898 & \textbf{0.931} &    & 0.915 & 0.962 & \textbf{0.967} &    & 0.990 & 0.950 \\
   &    &    & 0.8\bigstrut[b] & 0.975 & 0.983 & \textbf{0.984} &    & 0.996 & \textbf{0.997} & 0.997 &    & 0.990 & 0.950 \\
\cmidrule{2-14}   & \multirow{6}[2]{*}{100} & \multirow{3}[1]{*}{40} & 0.3 & 0.530 & 0.713 & \textbf{0.847} &    & 0.539 & 0.686 & \textbf{0.848} &    & 0.990 & 0.950 \\
   &    &    & 0.5 & 0.530 & 0.898 & \textbf{0.929} &    & 0.539 & 0.793 & \textbf{0.933} &    & 0.990 & 0.950 \\
   &    &    & 0.8\bigstrut[b] & 0.847 & \textbf{0.966} & 0.966 &    & \textbf{0.978} & 0.977 & 0.977 &    & 0.990 & 0.950 \\
   &    & \multirow{3}[1]{*}{10} & 0.3 & 0.530 & 0.713 & \textbf{0.844} &    & 0.539 & 0.685 & \textbf{0.846} &    & 0.990 & 0.950 \\
   &    &    & 0.5 & 0.530 & 0.746 & \textbf{0.883} &    & 0.539 & 0.702 & \textbf{0.905} &    & 0.990 & 0.950 \\
   &    &    & 0.8\bigstrut[b] & 0.531 & 0.926 & \textbf{0.933} &    & 0.564 & 0.944 & \textbf{0.944} &    & 0.990 & 0.950 \\
\bottomrule
\end{tabularx}%
\begin{tablenotes}
\item \textbf{Notes:} Results are based on $1,000$ Monte Carlo iterations. For each Monte Carlo iteration the corresponding sampling algorithms are run using $500$ draws, where the initial $500$ were discarded as burn-in. The values given for RMSE($\boldsymbol{\beta}$) and RMSE($\rho$) correspond to the average root mean squared error over all Monte Carlo iterations. Bold values denote the best performing specification within a section (symmetric or non-symmetric). The exogenous $\boldsymbol{\Omega}$ specifications correspond to classic SAR models with randomly perturbed exogenous adjacency matrices, which have an accuracy of 99\% and 95\% compared to the \emph{true} adjacency matrix. For RMSEs, lower values indicate outperformance. Conversely, for the accuracy indicators of $\boldsymbol{\Omega}$, higher values indicate outperformance.
\end{tablenotes}
  \label{tab:MC1}%
\end{threeparttable}%
\end{table}

\noindent

Table \ref{tab:MC1} summarizes the results of our Monte Carlo simulation. For all combinations of $N$, $T$, $\tilde{\rho}$ under scrutiny, the table presents the respective root mean square error for both the slope coefficients $\boldsymbol{\beta}$ and the spatial autoregressive parameter. The third block of the table shows the accuracy of the estimated adjacency matrix $\boldsymbol{\Omega}$. Lower values in terms of RMSEs indicate outperformance. Conversely, for accuracy in $\boldsymbol{\Omega}$ higher values indicate outperformance. The best performance among the three employed prior scenarios within a subgroup is highlighted in bold. In addition, the last two columns in Table \ref{tab:MC1} show the results for the benchmark SAR models using exogenous randomly perturbed adjacency matrices with accuracy fixed at the $99\%$ and the $95\%$ level, respectively. 

Intuitively, the precision of the estimation improves as the number of observations $NT$ increases in proportion to the number of unknown parameters.\footnote{The number of unknown parameters amounts to $N^2+T+q_0+2$ and $N(N-1)/2+N+T+q_0+2$ for non-symmetric and symmetric spatial weight matrices, respectively.} The results in Table \ref{tab:MC1} largely confirm this intuition.  The performance indicators for both $\rho$ and $\boldsymbol{\Omega}$ also clearly improve for high levels of spatial autocorrelation ($\rho = 0.8$). In scenarios where the number of unknown parameters is smaller than the number of observations our approach even manages to outperform both rather hard benchmarks using exogenous spatial weight matrices close to the true DGP. This relative outperformance appears particularly pronounced when the strength of spatial dependence $\rho$ is large. In these settings, symmetric specifications (which resemble the true DGP) even manage to produce accuracy in the adjacency matrix close to unity. 

Particularly interesting results appear in the most challenging Monte Carlo scenarios, where the number of unknown parameters is particularly large relative to the number of observations ($N = 100$ and $T = 10$). In these scenarios, the number of parameters to be estimated exceeds the number of observations by a factor of more than ten. In these cases, prior specifications without using shrinkage appear to fail estimating the underlying spatial structure by producing rather poor accuracy measures. However, when employing sparsity priors, the table reveals that our approach still manages to produce relatively accurate predictive results. In the existence of pronounced spatial autocorrelation, the sparsity specifications even manage to closely track the predictive performance of the rather tough exogenous benchmarks. 

Note that the symmetric specifications (where we impose $\omega_{ij}=\omega_{ji}$) typically outperform their non-symmetric counterparts due to their resemblance to the true DGP. However, for settings where the number of unknown parameters is smaller than the number of observations both scenarios track each other closely. Among the alternative prior specifications under scrutiny, the table shows rather similar results (no clear best specification emerges) in scenarios where $N$ is small relative to $T$. However, for particularly over-parametrized settings (high $N$ and low $T$) the proposed \textit{sparsity} priors particularly outperform the \textit{fixed} setups. Specifically, even in the scenario with $N=100$ and $T=10$, the sparsity priors still perform comparatively well.\footnote{Figure \ref{fig:traceplot_MCMC} in the appendix illustrates the convergence properties of a random Monte Carlo sample for the case of $N=20$ and $T=10$. This case was chosen as it is similar to the settings in the empirical applications.}

\section{Empirical illustration}

To illustrate our proposed approach using real data, we estimate spatial panel specifications based on country-specific daily infection rates in the very early phase of the coronavirus pandemic. We use the COVID-19 data set provided by the Johns Hopkins University (\citealt{dong2020interactive}). The database contains information on (official) daily infections for a large panel of countries around the globe. For the empirical illustration, we focus on the very beginning of the outbreak by using data from 17th of February to the 20th of April of 2020. 

The starting date of our sample marks the beginning of the pandemic in major countries, such that large parts of Asia, Europe and North America can be included.\footnote{Countries without any (official) infections in the starting period have been excluded from the sample. We moreover exclude India as a clear outlier from the sample due to its particular small (official) infection rates throughout the observation period.} The choice of the end date is motivated by the results of \cite{krisztin2020spatial}, where the degree of spatial dependence among infections rates becomes insignificant after the 20th April, when the majority of countries in the sample implemented lockdown policies. 

For the empirical application we use data for the following countries: Australia (AUS), Bahrain (BHR), Belgium (BEL), Canada (CAN), China (CHN), Finland (FIN), France (FRA), Germany (DEU), Iran (IRN), Iraq (IRQ), Israel (ISR), Italy (ITA), Japan (JPN), Kuwait (KWT), Lebanon (LBN), Malaysia (MYS), Oman (OMN), Republic of Korea (KOR), Russian Federation (RUS), Singapore (SGP), Spain (ESP), Sweden (SWE), Thailand (THA), United Arab Emirates (ARE), United Kingdom (GBR), United States (USA), and Viet Nam (VNM). 

By including a biweekly time lag, our resulting panel thus comprises $N=27$ countries across the globe for a period of $T=19$ days.\footnote{With a biweekly time lag, the dependent variable thus captures data from 2nd of April to the 20th of April ($T=19$). For a better comparison, we have fixed the time period captured by $\boldsymbol{y}_t$ for all alternative specifications. It is moreover worth noting that a notable earlier starting date would result in relatively few (cross-sectional) observations. However, our results, are rather robust when considering a longer time horizon.} We follow work by \cite{guliyev2020determining}, \cite{krisztin2020spatial}, or \cite{han2021quantifying}, among others, and use panel versions of a spatial growth specification for the country-specific COVID-19 infections:

\begin{equation}
\label{eq:empequation}
\boldsymbol{y}_t = \boldsymbol{\mu}+ \tau_t + \rho\boldsymbol{Wy}_{t-r}+ \boldsymbol{x}_{t-14}\beta+\boldsymbol{Z}_{t-14}\boldsymbol{\beta}_0+\boldsymbol{\varepsilon}_t,
\end{equation}
where $\boldsymbol{y}_t=\boldsymbol{x}_t-\boldsymbol{x}_{t-14}$, and $\boldsymbol{x}_t$ is an $N\times1$ vector comprising the (logged) daily number of official cases per 100,000 inhabitants per country for time period $t=1,...,T$.\footnote{The spatial growth regression in (\ref{eq:empequation}) may be alternatively specified in levels rather than in log-differences by setting $\boldsymbol{y}_t=\boldsymbol{x}_t$. Results using this alternative specification are very similar and are presented in the appendix.} $\boldsymbol{\mu}$ and $\tau_t$ represent fixed effects for the countries and the time periods, respectively. $\boldsymbol{W}$ denotes the spatial weight matrix with spatial autoregressive parameter $\rho$ as defined before. We again primarily focus on row-stochastic weight matrices. Results based on spatial weight matrices without row-standardization are presented in the appendix. 

We also consider alternative model specifications using contemporaneous as well as temporal lags of the spatial lag ($\boldsymbol{Wy}_{t-r}$ with $r\in\{0,14\}$). A plethora of recent studies exploit the contemporaneous spatial information ($r=0$) for modelling the spread of COVID-19 infections (among others, see \citealt{han2021quantifying}, \citealt{jaya2021bayesian}, \citealt{kosfeld2021covid}, \citealt{guliyev2020determining}, or  \citealt{krisztin2020spatial}). Using contemporaneous spatial information appears  reasonable when the primary interest lies in quantifying  spatial co-movements of  infection rates. However, for many questions of interest, a temporal spatial lag $\boldsymbol{Wy}_{t-r}$ ($r>0$) might be an interesting alternative since it reflects the notion that the spatial process of virus transmission takes some time to manifest (\citealt{elhorst2021dynamic}, \citealt{mitzekosfeld2021}). Since our proposed estimation approach can be easily applied to  these alternative specifications, we provide estimates for both specifications.\footnote{It is worth noting that in the special case of $r>0$, computational efficiency is tremendously increased, as no log-determinant calculations are required in the MCMC algorithm. The sampling strategy for these cases is presented in the appendix.}


In addition to the \emph{Initial infections} variable $\boldsymbol{x}_{t-14}$, matrix $\boldsymbol{Z}_{t-14}$ contains three explanatory variables on a daily basis. Several studies emphasize the importance of climatic condition on the COVID-19 virus spread. For a survey on the effects of climate on the spread of the COVID-19 pandemic, see \cite{briz2020effect}. We therefore use daily data on the country specific maximum measured temperature (\emph{Temperature}) and precipitation levels (\emph{Precipitation}) as additional covariates. Both variables stem from a daily database of country-specifc data, which was compiled via the Dark Sky API.\footnote{\url{https://www.kaggle.com/datasets/vishalvjoseph/weather-dataset-for-covid19-predictions}} As a third variable, we also include the well-known stringency index (\emph{Stringency}) put forward by \cite{hale2020variation}, which summarizes country-specific governmental policy measures to contain the spread of the virus. In this application, we use the biweekly average of the reported stringency index. Since all these influences arguably require some time to be reflected in the official infection figures, we use a biweekly lag of $14$ days (in accordance with $r$ in alternative variants).\footnote{As robustness checks, we have also tried a shorter lag length of one week. The estimated spatial structures appeared very similar to the biweekly benchmarks. All these additional robustness checks, along with the R codes, are available from the authors upon request.}

\begin{table}[ht]
  \centering
\begin{threeparttable}[tbp]

  \caption{Estimation results for benchmark specifications\label{tab:results}}
  \scriptsize
 \renewcommand{\arraystretch}{0.8}

\begin{tabular}{lrrrrrrrr}
\toprule
  & \multicolumn{4}{c}{$\boldsymbol{Wy}_{t}$} & \multicolumn{4}{c}{$\boldsymbol{Wy}_{t-14}$} \\
   & \multicolumn{2}{c}{Fixed} & \multicolumn{2}{c}{Sparsity} & \multicolumn{2}{c}{Fixed} & \multicolumn{2}{c}{Sparsity} \\
   & \multicolumn{1}{l}{Mean} & \multicolumn{1}{l}{Std.Dev.} & \multicolumn{1}{l}{Mean} & \multicolumn{1}{l}{Std.Dev.} & \multicolumn{1}{l}{Mean} & \multicolumn{1}{l}{Std.Dev.} & \multicolumn{1}{l}{Mean} & \multicolumn{1}{l}{Std.Dev.} \\
\midrule
Initial infections & \textbf{-0.8761} & 0.0117 & \textbf{-0.9244} & 0.0117 & \textbf{-0.9533} & 0.0126 & \textbf{-0.9911} & 0.0114 \\
Stringency & \textbf{-0.4566} & 0.0736 & \textbf{-0.5661} & 0.0451 & \textbf{-0.2503} & 0.0858 & 0.0616 & 0.0410 \\
Precipitation & 0.0365 & 0.0339 & -0.0444 & 0.0335 & 0.0541 & 0.0608 & 0.0483 & 0.0511 \\
Temperature & -0.0014 & 0.0015 & -0.0016 & 0.0015 & -0.0032 & 0.0026 & -0.0017 & 0.0025 \\
$\rho$ & \textbf{0.6319} & 0.0129 & \textbf{0.5592} & 0.0101 & \textbf{0.9618} & 0.0110 & \textbf{0.9481} & 0.0139 \\
$\sigma^2$ & \textbf{0.0187} & 0.0013 & \textbf{0.0209} & 0.0014 & \textbf{0.0401} & 0.0034 & \textbf{0.0516} & 0.0036 \\\midrule
Avg. \# neighbours & 7.8370 &    & 3.6083 &    & 4.2849 &    & 2.8082 &  \\
Fixed effects & \text{Yes} & & \text{Yes} & & \text{Yes} & & \text{Yes} & \\ 
$N$ & 27 & & 27 & & 27 & & 27 & \\
$T$ & 19 & & 19 & & 19 & & 19 & \\
\bottomrule
\end{tabular}%

\begin{tablenotes}
\scriptsize \item \textbf{Notes:} Posterior quantities based on $5,000$ MCMC draws, where the first $2,500$ were discarded as burn-ins. Values in bold denote significance under a 90\% posterior credible interval.
\end{tablenotes}
\end{threeparttable}%
\end{table}

Table \ref{tab:results} presents a summary of the estimation results. The left part of the table shows results for specifications using a contemporaneous spatial lag $\boldsymbol{Wy}_{t}$, while the right part summarizes results for the case $\boldsymbol{Wy}_{t-14}$.

For each specification, the first rows contain the posterior mean and standard deviations  for the slope parameters followed by estimates of $\rho$ and $\sigma^2$. Posterior quantities which appear significantly different from zero using a 90\% posterior credible interval are depicted in bold. The table moreover presents the average posterior expected number of neighbours, which is given by the average row sum of the matrix of posterior inclusion probabilities based on $p(\omega_{ij}=1|\mathcal{D})$. This measure can be viewed as a measure of sparsity in the estimated matrix of linkages. All specifications moreover contain fixed effects for both $N$ and $T$.\footnote{For the benchmark specifications, the number of unknown parameters and observations thus amounts to $753$ and $513$, respectively.}

Table \ref{tab:results} shows rather similar $\rho$ and $\sigma^2$ posterior quantities for the flat and the sparsity prior. However, there appear some marked differences between the specifications $\boldsymbol{Wy}_{t}$ and $\boldsymbol{Wy}_{t-14}$. In all cases, spatial dependence appears strong and precisely estimated, but  appears particularly high in the temporal lag specification $\boldsymbol{Wy}_{t-14}$. However, the table similarly reveals higher estimates for the nuisance parameter $\sigma^2$ for the temporal spatial lag models. The table shows rather precise and negative coefficients for the initial infections variable, indicating conditional convergence patterns. For most model variants the table moreover suggests a significant negative impact of the stringency index on infection growth. The majority of the slope parameter estimates associated with the variables temperature and precipitation appear more muted and insignificant. Overall, the table moreover clearly demonstrates that a hierarchical prior setup can enforce sparsity in the resulting adjacency matrix. Both sparsity specifications result in an average number of neighbours smaller than the models with fixed prior specifications.   

\begin{figure}[!ht]
\caption{Posterior inclusion probabilities for benchmark specifications}
\label{fig:pips}
\centering
\begin{minipage}[b]{.45\linewidth}
\subcaption*{Fixed ($\underline{p} = 1/2$); $\boldsymbol{Wy}_t$}\vspace{-0.75cm}\label{fig:2a}
\centering \includegraphics[scale=0.35]{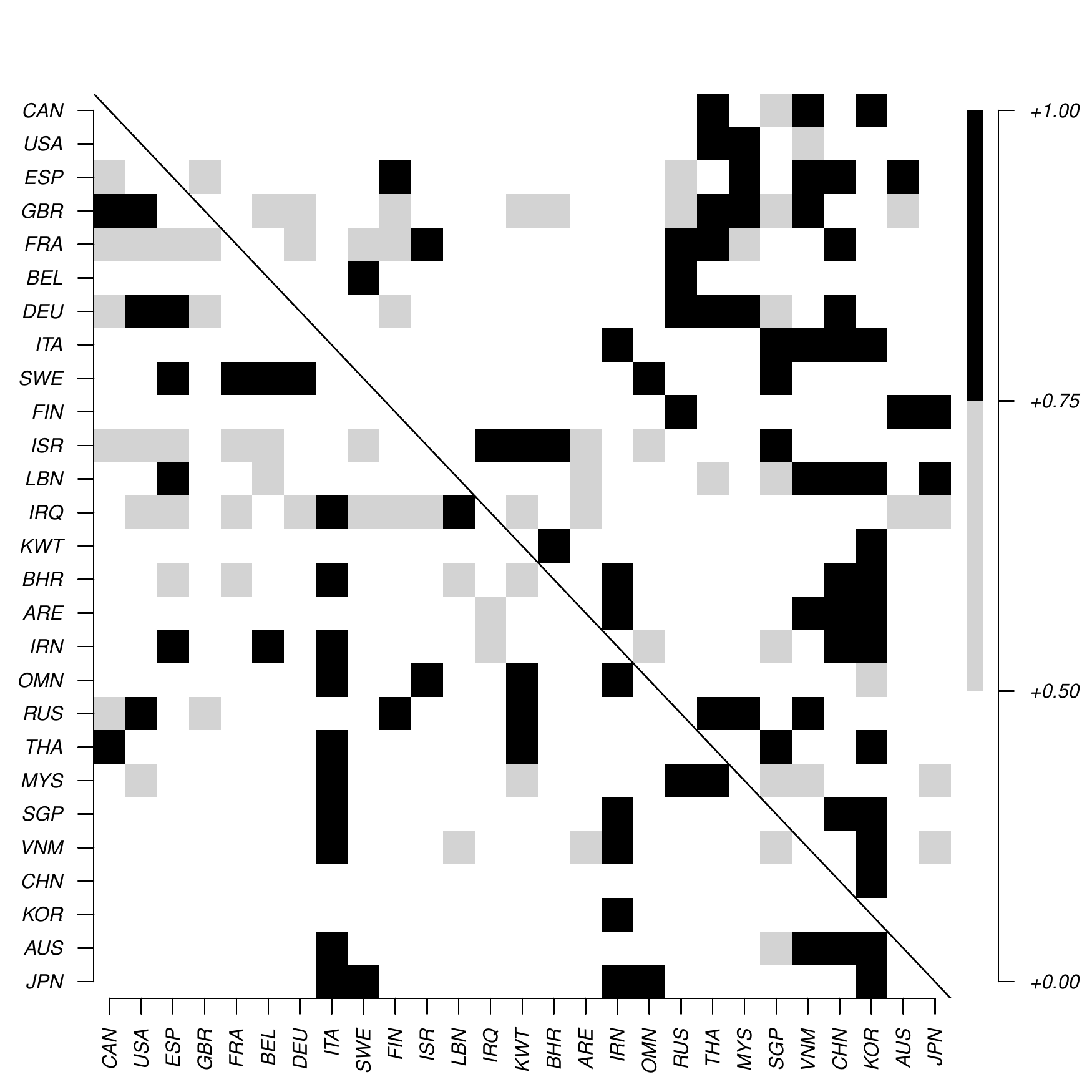}
\end{minipage}%
\begin{minipage}[b]{.45\linewidth}
\subcaption*{Sparsity ($\underline{m} = 7$); $\boldsymbol{Wy}_t$}\vspace{-0.75cm}\label{fig:2b}
\centering \includegraphics[scale=0.35]{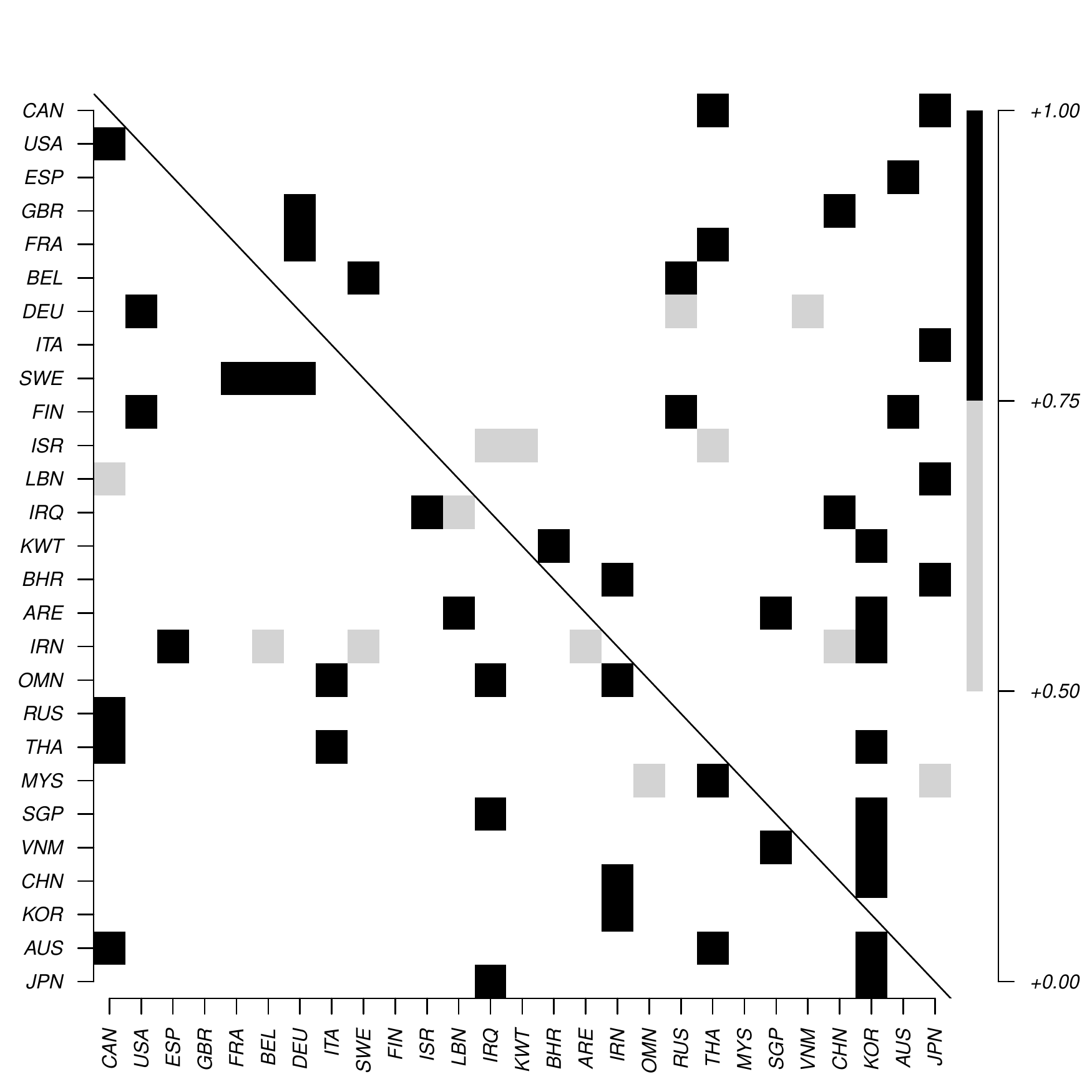}
\end{minipage}\\ \vspace{0.5cm}
\begin{minipage}[b]{.45\linewidth}
\subcaption*{Fixed ($\underline{p} = 1/2$); $\boldsymbol{Wy}_{t-14}$}\vspace{-0.75cm}\label{fig:2c}
\centering \includegraphics[scale=0.35]{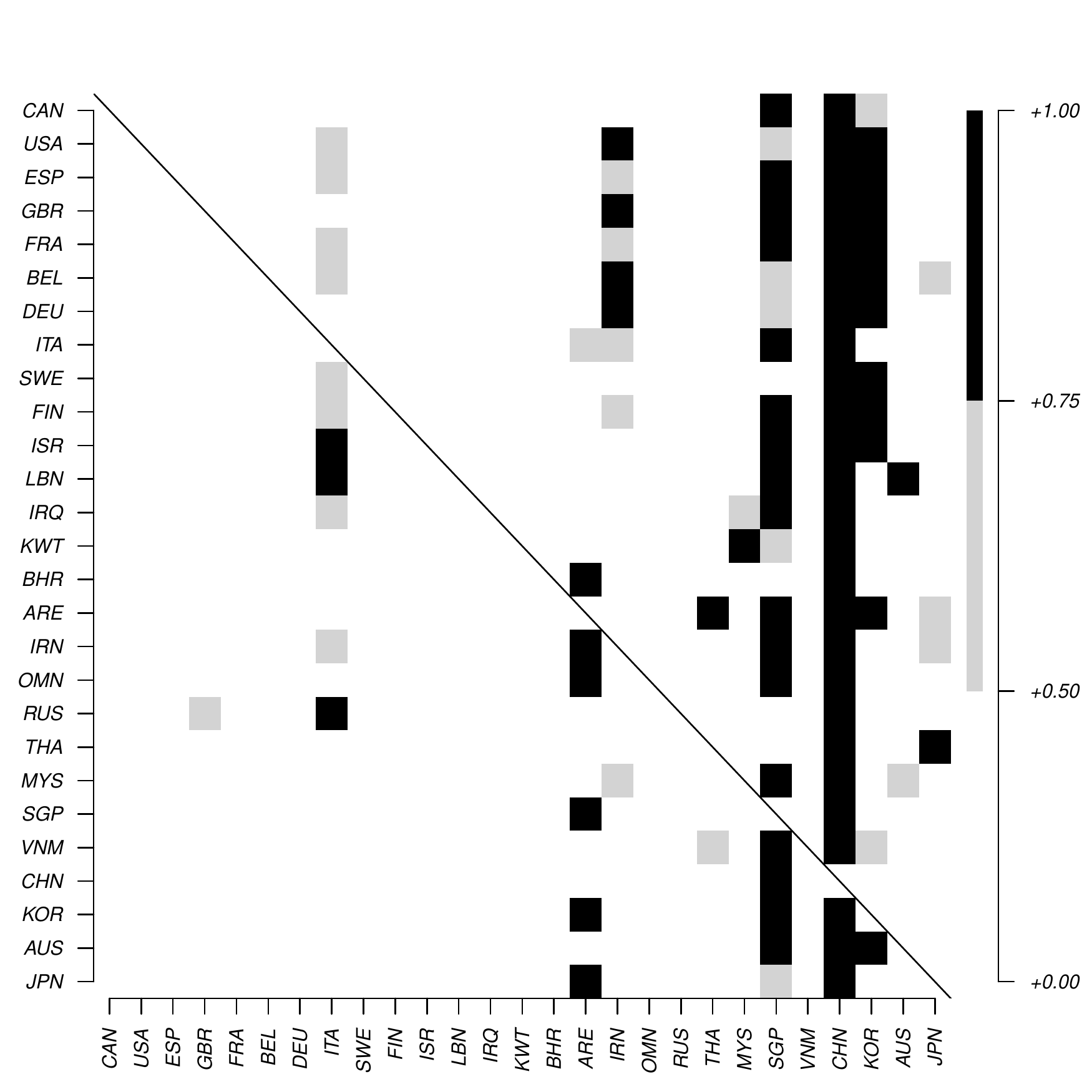}
\end{minipage}
\begin{minipage}[b]{.45\linewidth}
\subcaption*{Sparsity ($\underline{m} = 7$); $\boldsymbol{Wy}_{t-14}$}\vspace{-0.75cm}\label{fig:2d}
\centering \includegraphics[scale=0.35]{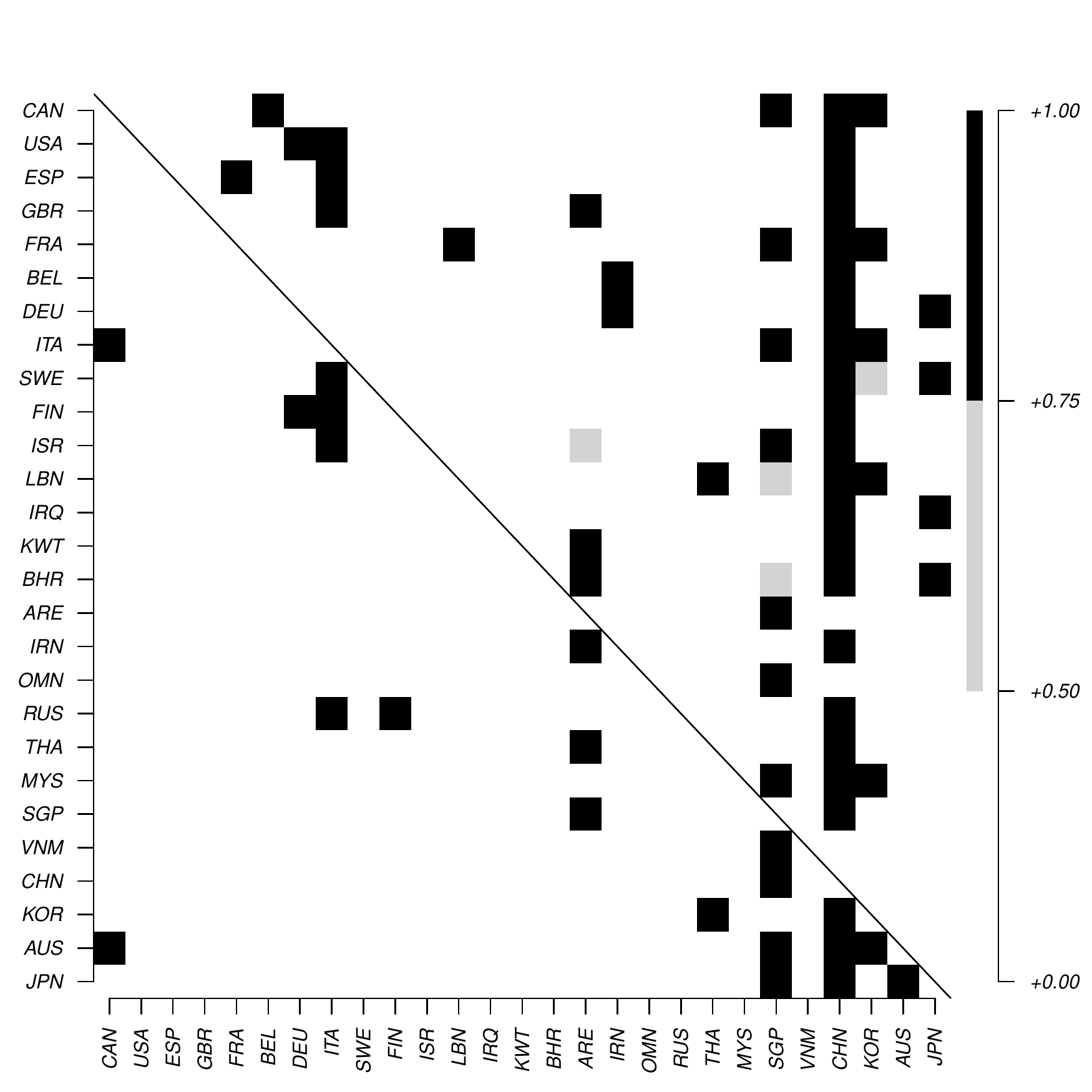}
\end{minipage}\\
\begin{minipage}{12.5cm}~\\
\scriptsize \textbf{Notes}: Posterior inclusion probabilities of spatial links based on $5,000$ MCMC draws. Inclusion probabilities 0.50-0.75 (little evidence for inclusion) are coloured grey. Strong evidence for inclusion (>0.75) indicated by black colour. 
\end{minipage}%
\end{figure}

Figure $\ref{fig:pips}$ depicts the posterior inclusion probabilities $p(\omega_{ij}=1|\mathcal{D})$ for the considered specifications. To better visualize the results we have reordered the countries by their longitudes, starting with Canada and the United States and ending with south-east Asian countries, Australia and Japan. Clusters along the main diagonal thus roughly indicate geographic spatial linkages.  For the sake of visualization, we distinguish between negligible evidence for inclusion ($<0.50$; white colour), moderate evidence ($0.50-0.75$; grey colour), and strong evidence ($>0.75$; black colour).

The two upper plots in Figure \ref{fig:pips} depict posterior inclusion probabilities $p(\omega_{ij}=1|\mathcal{D})$ for the specifications involving a contemporaneous spatial lag $\boldsymbol{Wy}_t$, while the lower part shows temporal spatial lag specifications $\boldsymbol{Wy}_{t-14}$. In both cases, the left subplots present results based on independent prior inclusion probabilities of $\underline{p}=1/2$. The right plots are based on sparsity priors using $\underline{m}=7$. The columns in the subplots indicate marginal posterior importance of the countries as predictors of coronavirus infections in linked countries. Conversely, rows depict the countries to be predicted. The results using sparsity priors generally produce similar patterns as the fixed prior specifications and clearly demonstrate its ability of dimension reduction in the connectivity structure. For the contemporaneous spatial lag specification (upper plots), the figure suggests a slightly more pronounced regional dependency structure as compared to the temporal spatial lags. The figure moreover reveals marked spill-out effects from Asian countries, as well as from Iran and Italy.\footnote{The regional dependency structure appears particularly pronounced when a level specification of the infection dynamics is imposed. Sensitivity checks based on this alternative specifications are presented in Figure \ref{fig:pips_level_spec} in the appendix.}

Results based on a biweekly temporal spatial lag $\boldsymbol{Wy}_{t-14}$ show even more pronounced spill-out effects from Asian countries (most notably China, Republic of Korea, and Singapore).\footnote{When comparing the results, it is important to note that for all specifications under scrutiny, we have fixed time period in the dependent variable ($\boldsymbol{y}_t$ ranges from the 2nd February to the 20th of February; i.e. $T=19$).The biweekly temporal spatial lag specification thus inherently comprises spatial information prior to the period in $\boldsymbol{y}_t$.} For European countries, results similarly suggest Italy as a further important source country of spatial virus transmission. The estimated spatial linkages are thus in close agreement with the actual origins of the overall virus transmission for the very early period of the global outbreak of the pandemic.

\begin{figure}[!ht]
\caption{Trace plots for benchmark specifications}
\label{fig:trace}
\centering
\begin{minipage}[b]{.5\linewidth}
\centering \includegraphics[scale=.77]{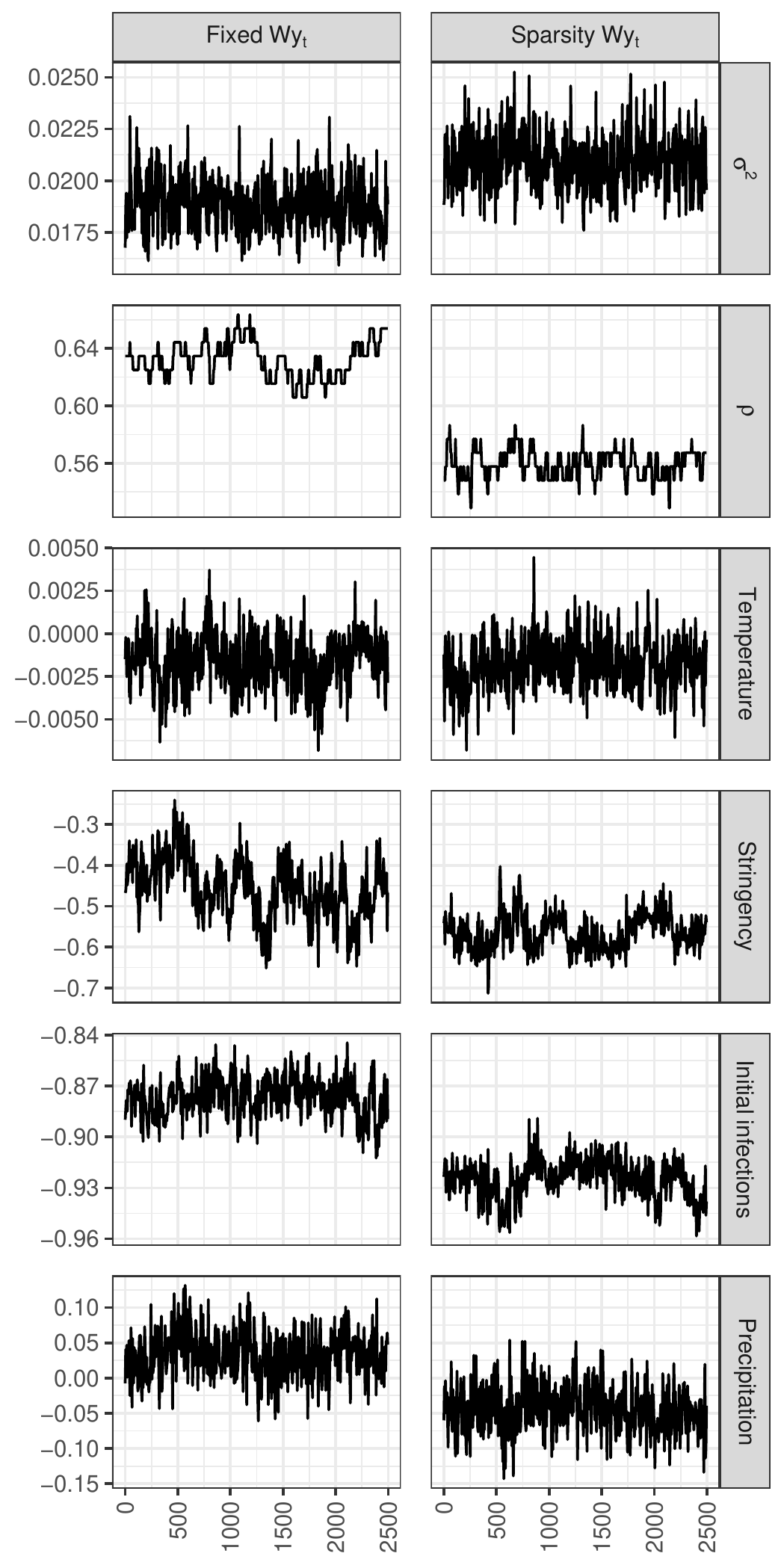}
\end{minipage}%
\begin{minipage}[b]{.5\linewidth}
\centering \includegraphics[scale=.77]{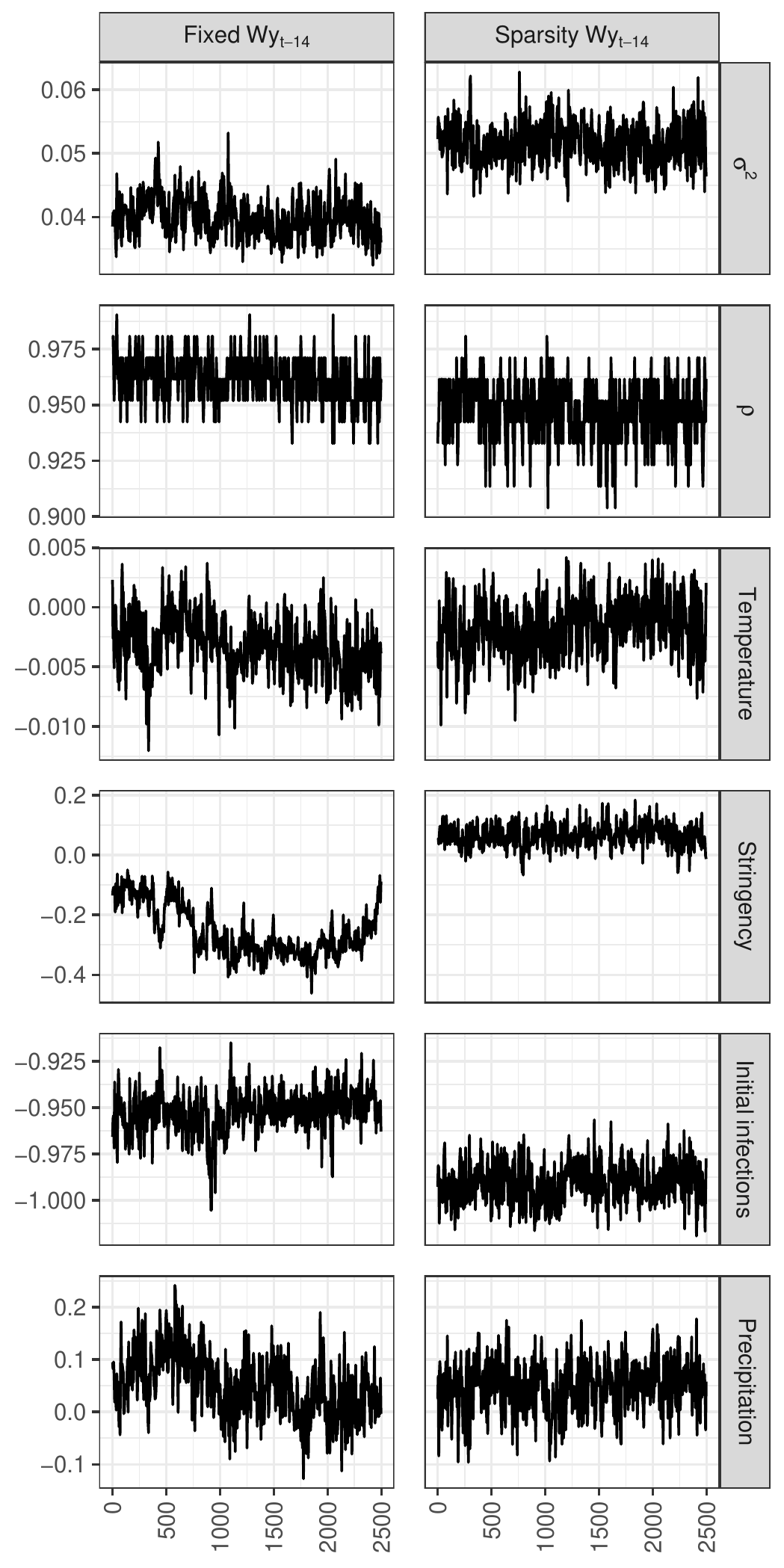}
\end{minipage}\\ 
\begin{minipage}{\linewidth}
\scriptsize \textbf{Notes}: Posterior draws based on $5,000$ MCMC draws, where the first $2,500$ were discarded as burn-ins. 
\end{minipage}%
\end{figure}

To showcase convergence of the posterior MCMC chains, Figure \ref{fig:trace} depicts trace plots for $\rho$, $\sigma^2$, and slope parameters. Overall, the trace plots show rather good mixing and convergence properties. Convergence of the chains have moreover been checked using the diagnostics proposed by \cite{geweke1992evaluating} implemented in the R package \textbf{coda} (\citealt{plummer2006coda}). Results moreover appear rather robust concerning alternative modelling frameworks. Estimation results of these alternative specifications are presented in the appendix.\footnote{Estimates when using a smaller time lag of seven days also appear very similar. Results along with the R codes used are available from the authors upon request.}

\section{Concluding remarks}

In this paper we propose a Bayesian approach for estimation of weight matrices in spatial econometric models. A particular advantage of our approach is the simple integration into a standard Bayesian  MCMC algorithm. The proposed framework can therefore be adapted and extended in a simple and computationally efficient way to cover a large number of alternative spatial specifications prevalent in recent literature. Our approach may thus be easily extended to cover inter alia non-Gaussian models such as spatial probit \citep{lesage2011new} or logit specifications \citep{Krisztin2019}, local spillover models \citep{HalleckVega2015}, or spatial error models \citep{LeSage2009}. 

Our approach does not not necessarily rely on specific prior information for the spatial linkages. Spatial information, however, can be easily implemented in a flexible and transparent way. We moreover motivate the use of hierarchical priors which impose sparsity in the resulting spatial weight matrix. These sparsity priors are particularly useful in applications where the number of unknown parameters exceeds those of the observations. The virtues of our approach comes at the price that we focus on spatial neighbourhood structures which are binary (prior to row-standardization). However, this assumption is implicitly assumed in many spatial applications in the regional economic literature where spatial weight matrices are constructed based on concepts of contiguity, distance band, or nearest neighbours.  

Based on Monte Carlo simulations, we show that our approach appears particularly promising when the number of spatial observations $N$ is large relative to the time dimension $T$, which is a rather common characteristic of data sets in the regional science literature. We moreover demonstrate the usefulness of our approach using real data on the outbreak of the COVID-19 pandemic. The results of this empirical application corroborate the findings in the Monte Carlo simulation study that the proposed approach performs well even in the cases of high over-parametrization. 


\newpage

\appendix

\renewcommand\thesection{\Alph{section}}
\setcounter{section}{1}

\section*{Appendix}
\renewcommand{\thetable}{A\arabic{table}}
\renewcommand{\thefigure}{A\arabic{figure}}
\setcounter{figure}{0}
\setcounter{table}{0}

\subsection*{Estimation strategies for alternative spatial lag specifications}
In the empirical application, the paper also considers model variants with a spatial lag on the temporal lag of the dependent variable. The considered specification can be written as:
\begin{equation}
\label{eq:SARtlagapp}
\boldsymbol{y}_t= \boldsymbol{\mu}+\tau_t+\rho\boldsymbol{Wy}_{t-1}+\boldsymbol{Z}_t\boldsymbol{\beta}_0+\boldsymbol{\varepsilon}_t, \hspace{2cm}t=1,...,T,
\end{equation}
where $\boldsymbol{y}_{t-1}$ now denotes the temporal lag of the dependent variable and the other quantities are defined as before. From a Bayesian perspective, it is worth noting that an additional temporal lag of the dependent variable $\boldsymbol{y}_{t-1}$ can be treated like any other explanatory variable and thus part of the matrix of covariates $\boldsymbol{Z}_t$. 

From a computational perspective, the specification in Eq. (\ref{eq:SARtlagapp}) is much easier to deal with as compared to SAR models involving a contemporaneous spatial lag in the dependent variable (i. e. $\rho\boldsymbol{Wy}_t$). This is due to the fact that the likelihood function does not involve a determinant term.

To maintain succinct notation, we again collect the fixed effects along with the explanatory variables in a $N\times q$ matrix $\boldsymbol{X}_t$ and stack the quantities as before $\boldsymbol{X}=\left[\boldsymbol{X}_1', \dots, \boldsymbol{X}_T'\right]'$, with $\boldsymbol{Y}_t$ and $\boldsymbol{Y}_{t-1}$ denoting the stacked $NT\times 1$ vectors of the dependent variable and the lag, respectively.  Defining $\boldsymbol{e}_t=\boldsymbol{Y}_t-\rho(\boldsymbol{I}_T\otimes\boldsymbol{W})\boldsymbol{Y}_{t-1}-\boldsymbol{X\beta}$, the likelihood reduces to a much simpler form and is given by:
\begin{equation}
p(\mathcal{D}|\bullet)= \frac{1}{(2\pi\sigma^2)^{NT}} \exp\left[-\frac{1}{2\sigma^2}\boldsymbol{e}_t'\boldsymbol{e}_t\right]. \label{eq:SARtlaglikapp}
\end{equation}

By using the same prior specifications like in the SAR case, the posterior probabilities of including or excluding $\omega_{ij}$ conditional on the other parameters are then given by:
\begin{eqnarray}
p(\omega_{ij}=1|\boldsymbol{\Omega}_{-ij},\boldsymbol{\beta},\sigma^2,\rho,\mathcal{D})\propto p(\omega_{ij}=1)\exp\left[-\frac{1}{2\sigma^2}\boldsymbol{e}_1'\boldsymbol{e}_1\right],\\
p(\omega_{ij}=0|\boldsymbol{\Omega}_{-ij},\boldsymbol{\beta},\sigma^2,\rho,\mathcal{D})\propto p(\omega_{ij}=0)\exp\left[-\frac{1}{2\sigma^2}\boldsymbol{e}_0'\boldsymbol{e}_0\right],
\end{eqnarray}
where $\boldsymbol{e}_1$ and $\boldsymbol{e}_0$ denote the updated vector of residuals $\boldsymbol{e}$ when $\omega_{ij}=1$ and $\omega_{ij}=0$, respectively. The conditional Bernoulli posterior for $\omega_{ij}$ is given by:
\begin{equation}
p(\omega_{ij}|\boldsymbol{\Omega}_{-ij},\boldsymbol{\beta},\sigma^2,\rho,\mathcal{D})\sim\mathcal{BER}\left(\frac{\bar{p}_{ij}^{(1)}}{\bar{p}_{ij}^{(0)}+\bar{p}_{ij}^{(1)}}\right),
\end{equation}
with $\bar{p}_{ij}^{(1)}=p(\omega_{ij}=1|\boldsymbol{\Omega}_{-ij},\boldsymbol{\beta},\sigma^2,\rho,\mathcal{D})$ and $\bar{p}_{ij}^{(0)}=p(\omega_{ij}=0|\boldsymbol{\Omega}_{-ij},\boldsymbol{\beta},\sigma^2,\rho,\mathcal{D})$.

The remaining conditional posterior distributions required for the MCMC sampler are given by:
\begin{eqnarray}
p(\boldsymbol{\beta}|\sigma^2,\rho,\boldsymbol{\Omega},\mathcal{D}) &\sim& \mathcal{N}(\bar{\boldsymbol{b}}_\beta,\bar{\boldsymbol{V}}_\beta)\\
\bar{\boldsymbol{b}}_\beta&=&\sigma^{-2}\bar{\boldsymbol{V}}_\beta\boldsymbol{X}'[\boldsymbol{Y}-\rho(\boldsymbol{I}_T\otimes\boldsymbol{W})\boldsymbol{Y}_{t-1}] \notag \\
\bar{\boldsymbol{V}}_\beta&=&\left(\sigma^{-2}\boldsymbol{X}'\boldsymbol{X}+\underline{\boldsymbol{V}}_\beta^{-1}\right)^{-1}. \notag
\end{eqnarray}
\begin{eqnarray}
p(\sigma^2 | \boldsymbol{\beta},\rho,\boldsymbol{\Omega},\mathcal{D}) &\sim& \mathcal{IG}(\bar{a}_{\sigma^2},\bar{b}_{\sigma^2})\\
\bar{a}_{\sigma^2}&=&\underline{a}_{\sigma^2}+NT/2 \notag\\
\bar{b}_{\sigma^2}&=&\underline{b}_{\sigma^2}+\boldsymbol{e}_t'\boldsymbol{e}_t. \notag
\end{eqnarray}
Unlike the other parameters, the conditional posterior for $\rho$ again takes no well-known form and can be sampled by using a griddy-Gibbs or tuned Metropolis-Hastings step:
\begin{equation}
p(\rho|\boldsymbol{\beta},\sigma^2,\boldsymbol{\Omega},\mathcal{D})\propto p(\rho)\exp\left[-\frac{1}{2\sigma^2}\boldsymbol{e}_t'\boldsymbol{e}_t \right].
\end{equation}
When using a normal prior distribution for $p(\rho)$, it is worth noting that the spatial lag can be simply captured in the matrix of explanatory variables, such that the parameter $\rho$ is incorporated in the vector $\boldsymbol{\beta}$. However, in order to pay particular attention to model stability as well as prior consistency to the benchmark SAR specification in the main body of the paper, we similarly employ a beta prior for $\rho$, which results in the non-standard form of the conditional posterior for $\rho$.\footnote{When considering specifications with a spatial lag in the explanatory variables (typically referred to as SLX models), the MCMC sampling scheme is rather similar, which also considerably reduces the computational burden as compared to SAR frameworks.}

\subsection*{Empirical results for additional model specifications and Monte Carlo diagnostics}

\begin{figure}[!ht]
\caption{Posterior inclusion probabilities of linkages for level specifications}
\label{fig:pips_level_spec}
\centering
\begin{minipage}[b]{.45\linewidth}
\subcaption*{Fixed ($\underline{p} = 1/2$); $\boldsymbol{Wy}_t$}\vspace{-0.75cm}\label{fig:2a}
\centering \includegraphics[scale=0.35]{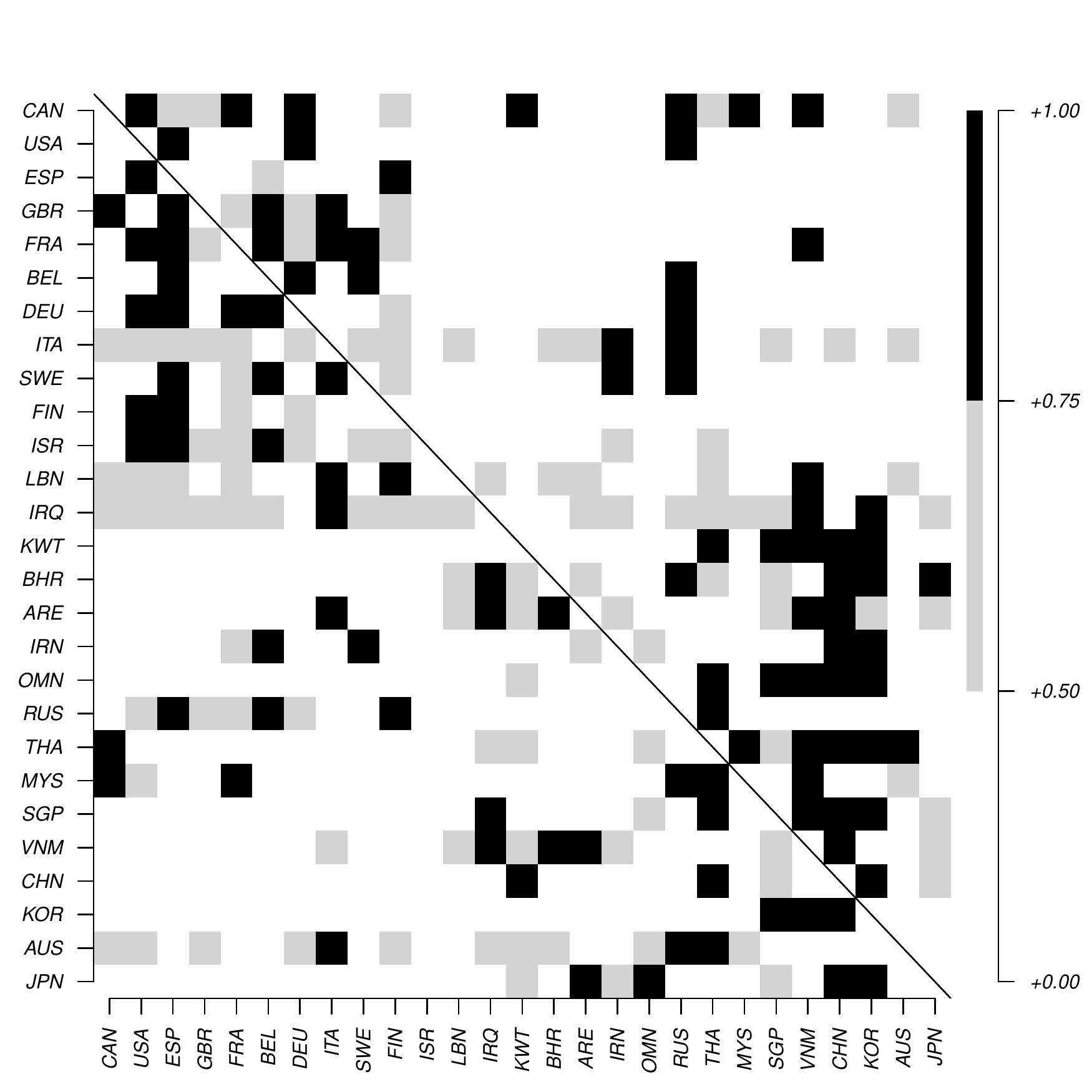}
\end{minipage}%
\begin{minipage}[b]{.45\linewidth}
\subcaption*{Sparsity ($\underline{m} = 7$); $\boldsymbol{Wy}_t$}\vspace{-0.75cm}\label{fig:2b}
\centering \includegraphics[scale=0.35]{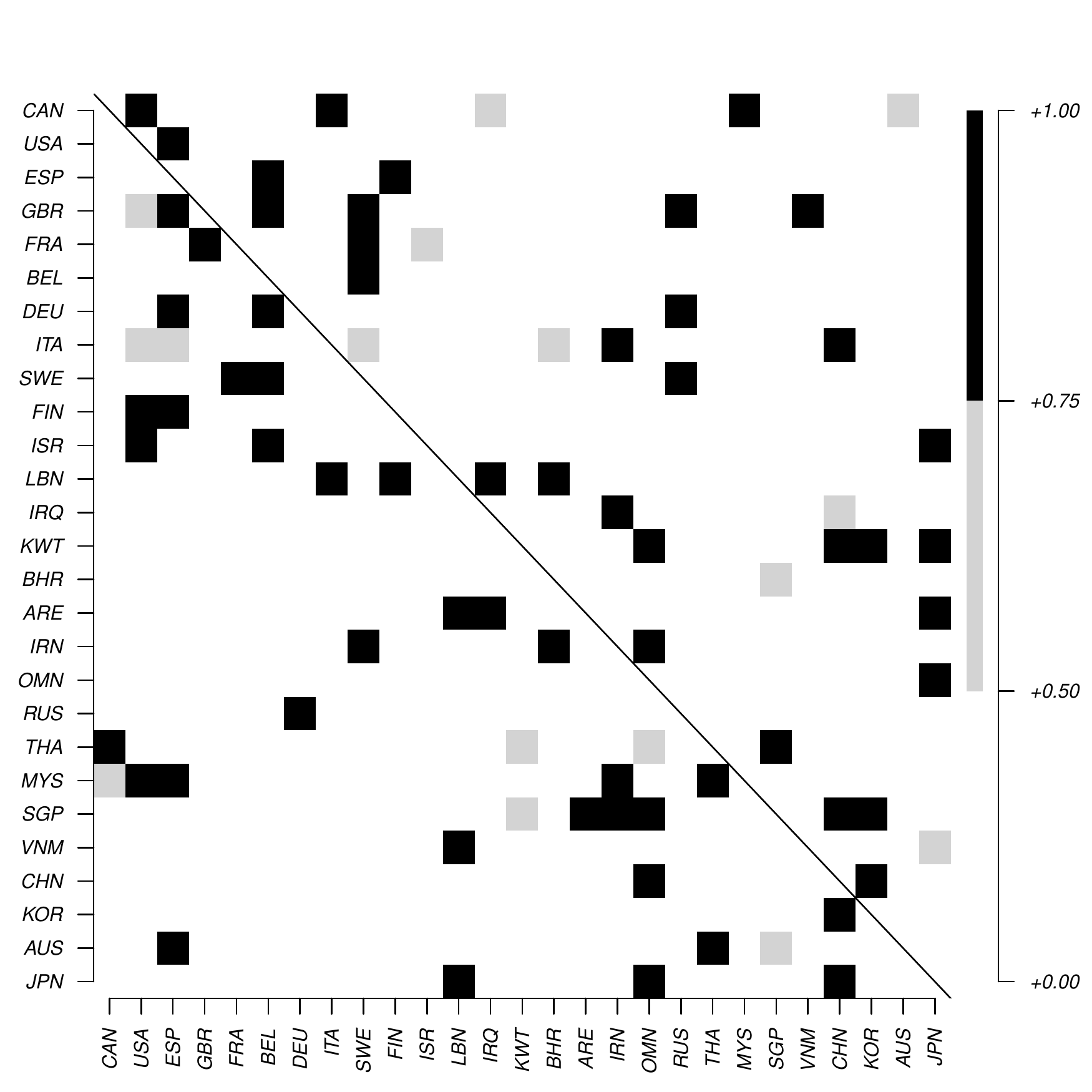}
\end{minipage}\\ \vspace{0.5cm}
\begin{minipage}[b]{.45\linewidth}
\subcaption*{Fixed ($\underline{p} = 1/2$); $\boldsymbol{Wy}_{t-14}$}\vspace{-0.75cm}\label{fig:2c}
\centering \includegraphics[scale=0.35]{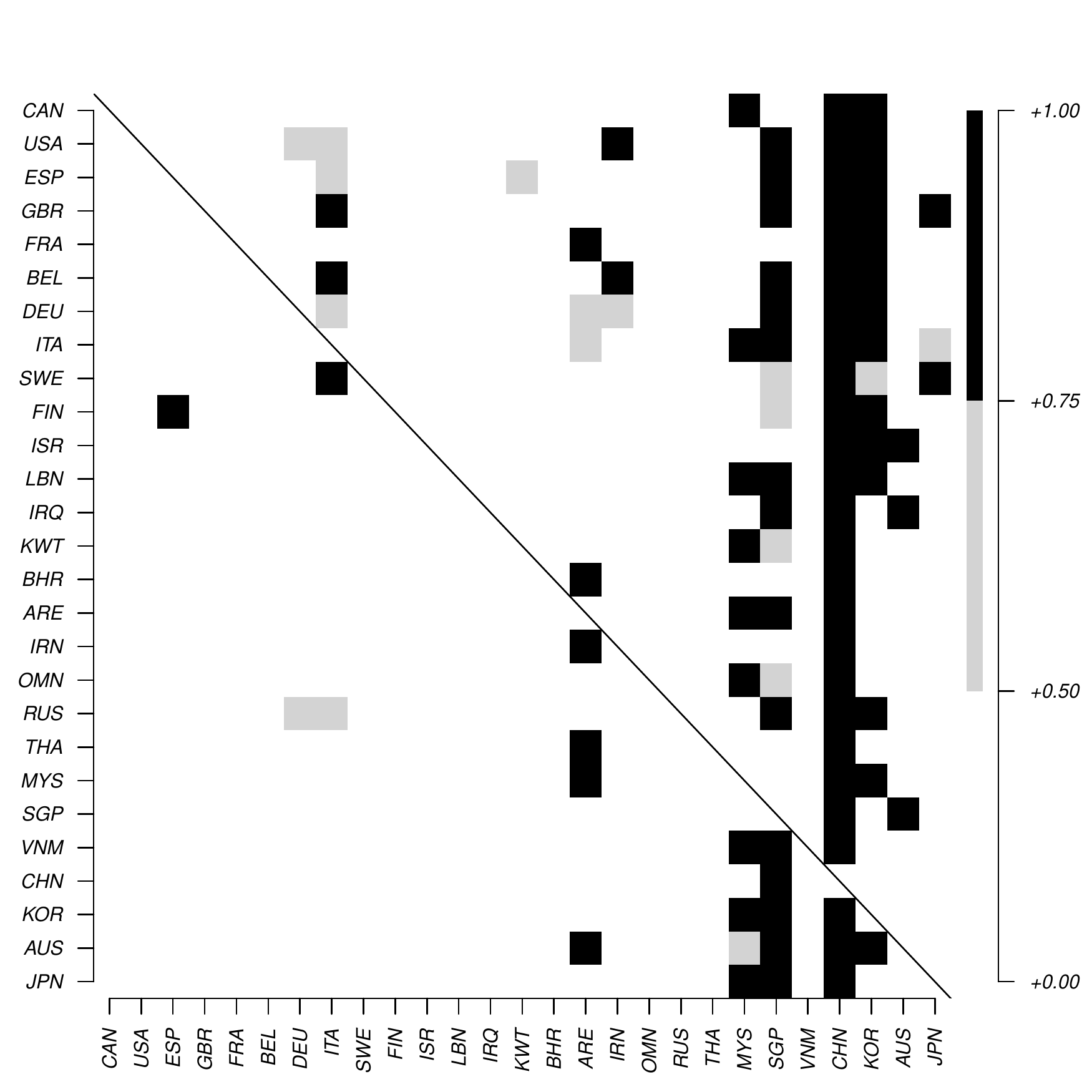}
\end{minipage}
\begin{minipage}[b]{.45\linewidth}
\subcaption*{Sparsity ($\underline{m} = 7$); $\boldsymbol{Wy}_{t-14}$}\vspace{-0.75cm}\label{fig:2d}
\centering \includegraphics[scale=0.35]{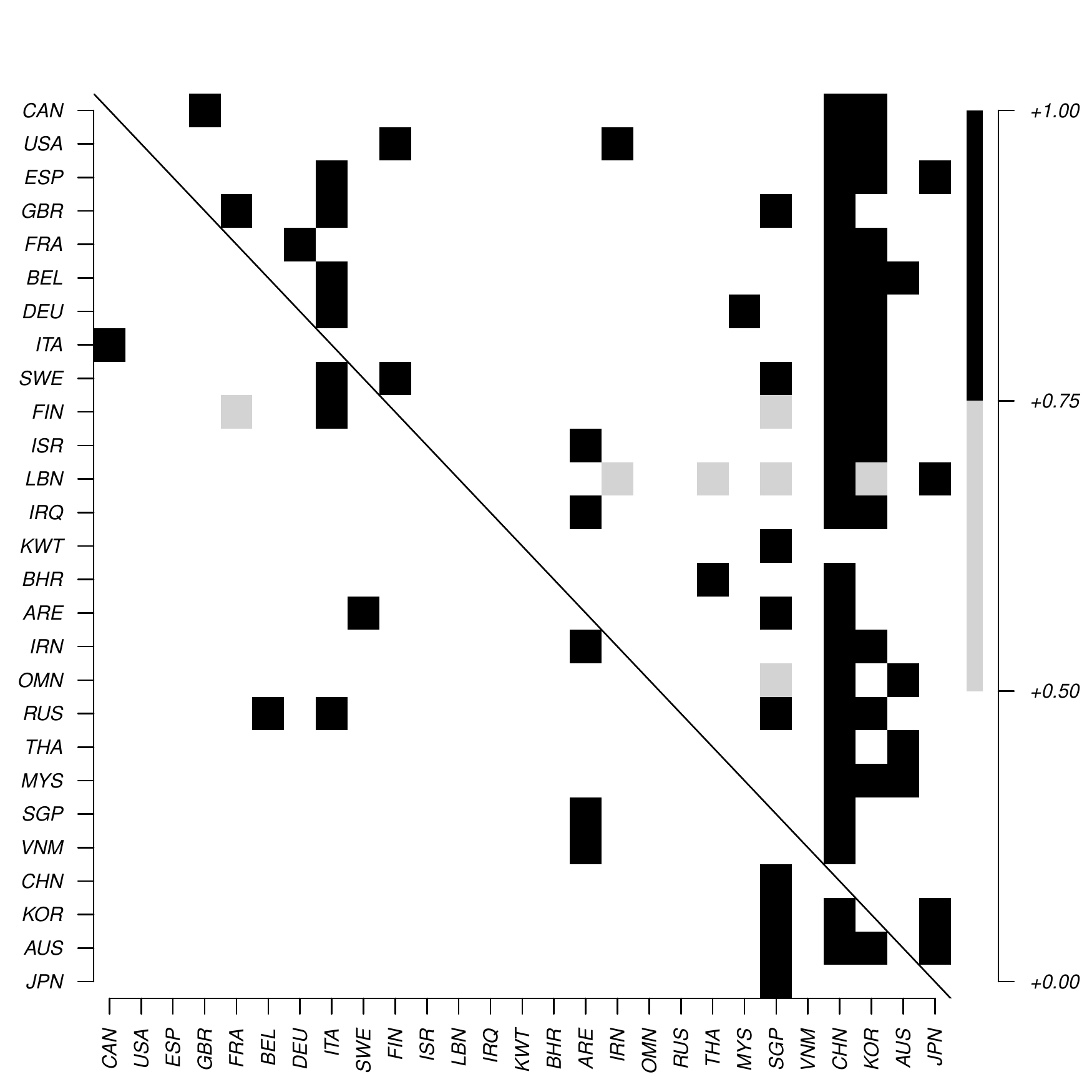}
\end{minipage}\\
\begin{minipage}{12.5cm}~\\
\scriptsize \textbf{Notes}: Posterior inclusion probabilities of spatial links based on $5,000$ MCMC draws. Inclusion probabilities 0.50-0.75 (little evidence for inclusion) are coloured grey. Strong evidence for inclusion (>0.75) indicated by black colour. Level specifications refer to specifications by using (all else being equal) log levels of infection rates rather than log-differences as the dependent variable.
\end{minipage}%
\end{figure}

\begin{figure}[!ht]
\caption{Posterior inclusion probabilities of linkages for specifications without row-standardization of the weight matrix}
\label{fig:pips_notstandardized}
\centering
\begin{minipage}[b]{.45\linewidth}
\subcaption*{Fixed ($\underline{p} = 1/2$); $\boldsymbol{Wy}_t$}\vspace{-0.75cm}\label{fig:2a}
\centering \includegraphics[scale=0.35]{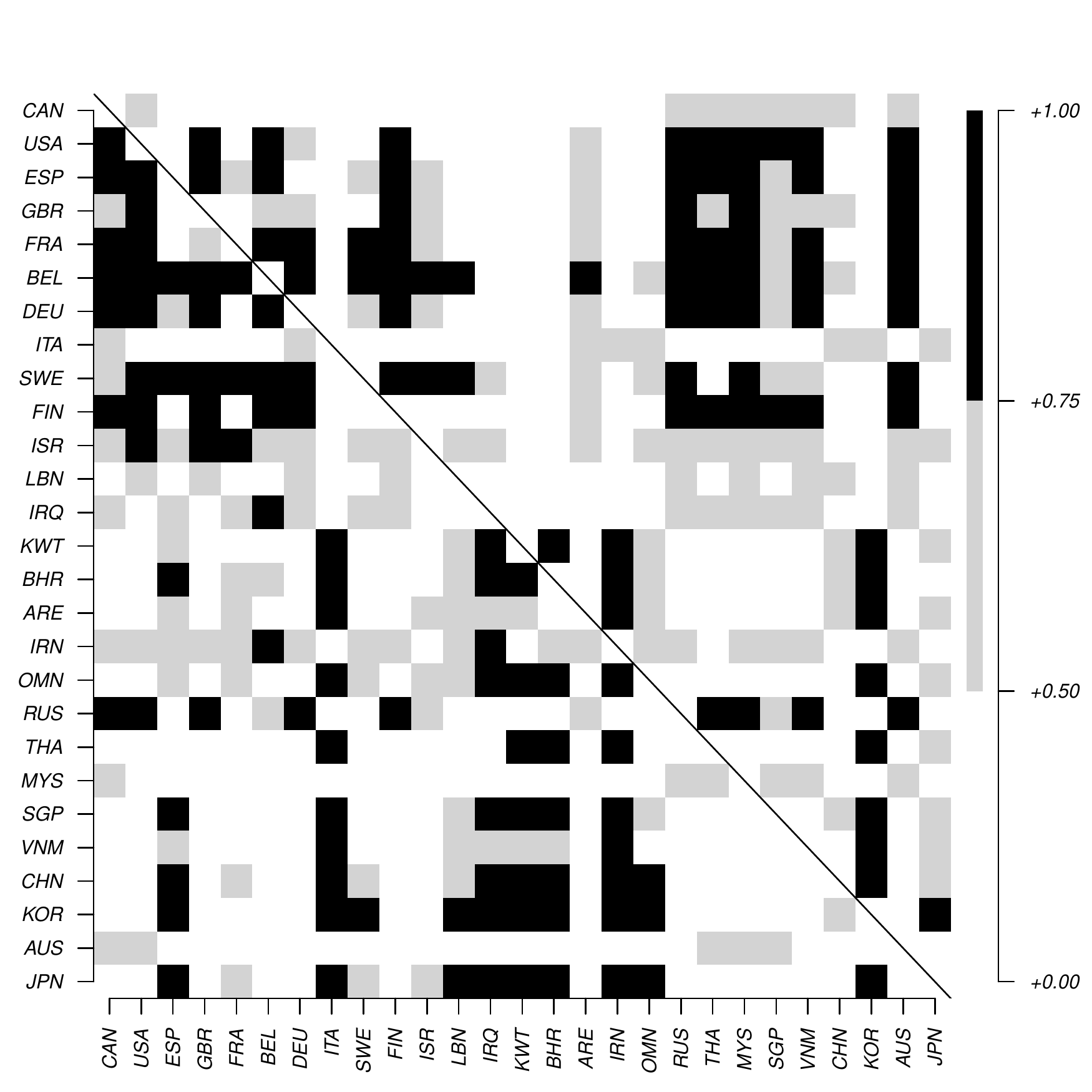}
\end{minipage}%
\begin{minipage}[b]{.45\linewidth}
\subcaption*{Sparsity ($\underline{m} = 7$); $\boldsymbol{Wy}_t$}\vspace{-0.75cm}\label{fig:2b}
\centering \includegraphics[scale=0.35]{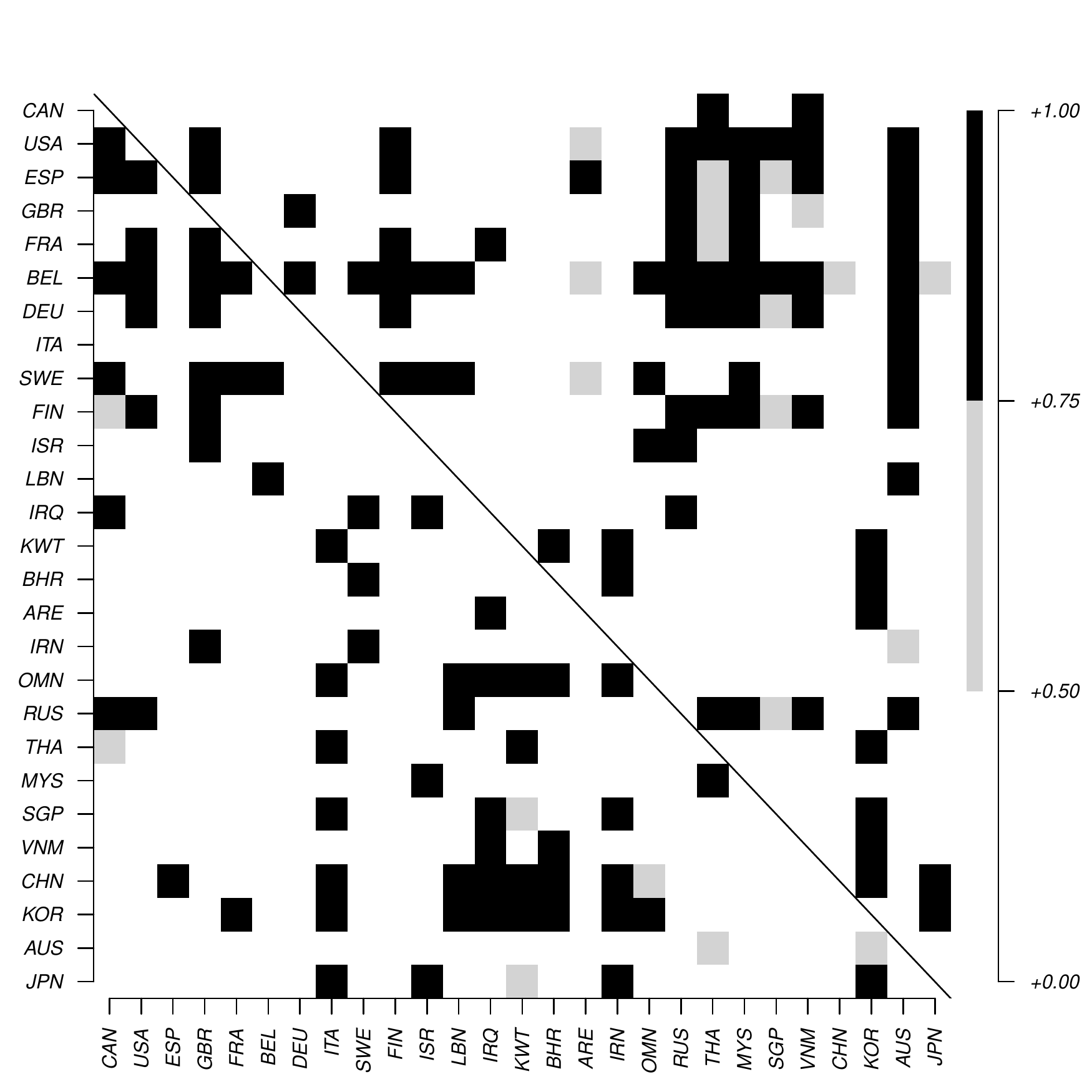}
\end{minipage}\\ \vspace{0.5cm}
\begin{minipage}[b]{.45\linewidth}
\subcaption*{Fixed ($\underline{p} = 1/2$); $\boldsymbol{Wy}_{t-14}$}\vspace{-0.75cm}\label{fig:2c}
\centering \includegraphics[scale=0.35]{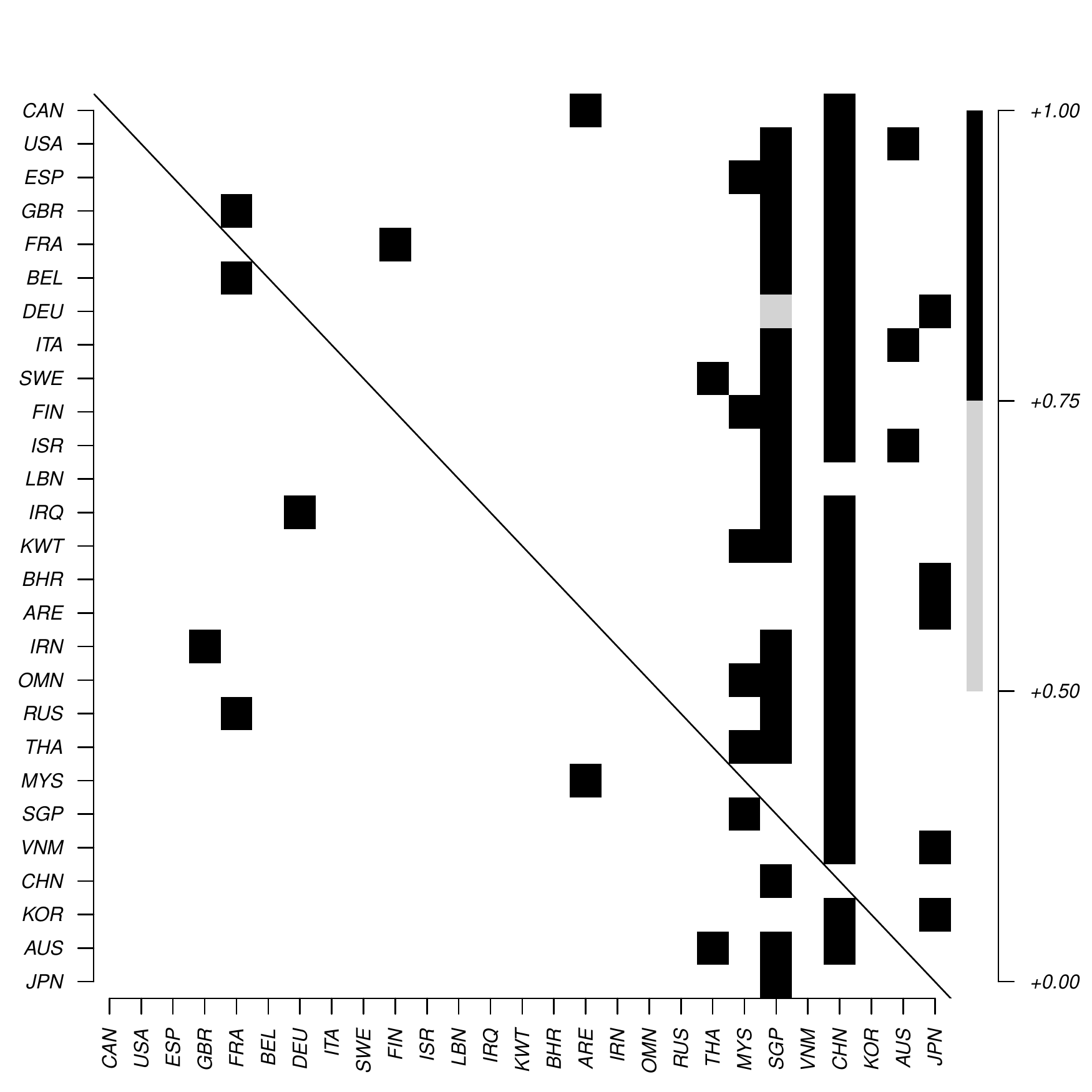}
\end{minipage}
\begin{minipage}[b]{.45\linewidth}
\subcaption*{Sparsity ($\underline{m} = 7$); $\boldsymbol{Wy}_{t-14}$}\vspace{-0.75cm}\label{fig:2d}
\centering \includegraphics[scale=0.35]{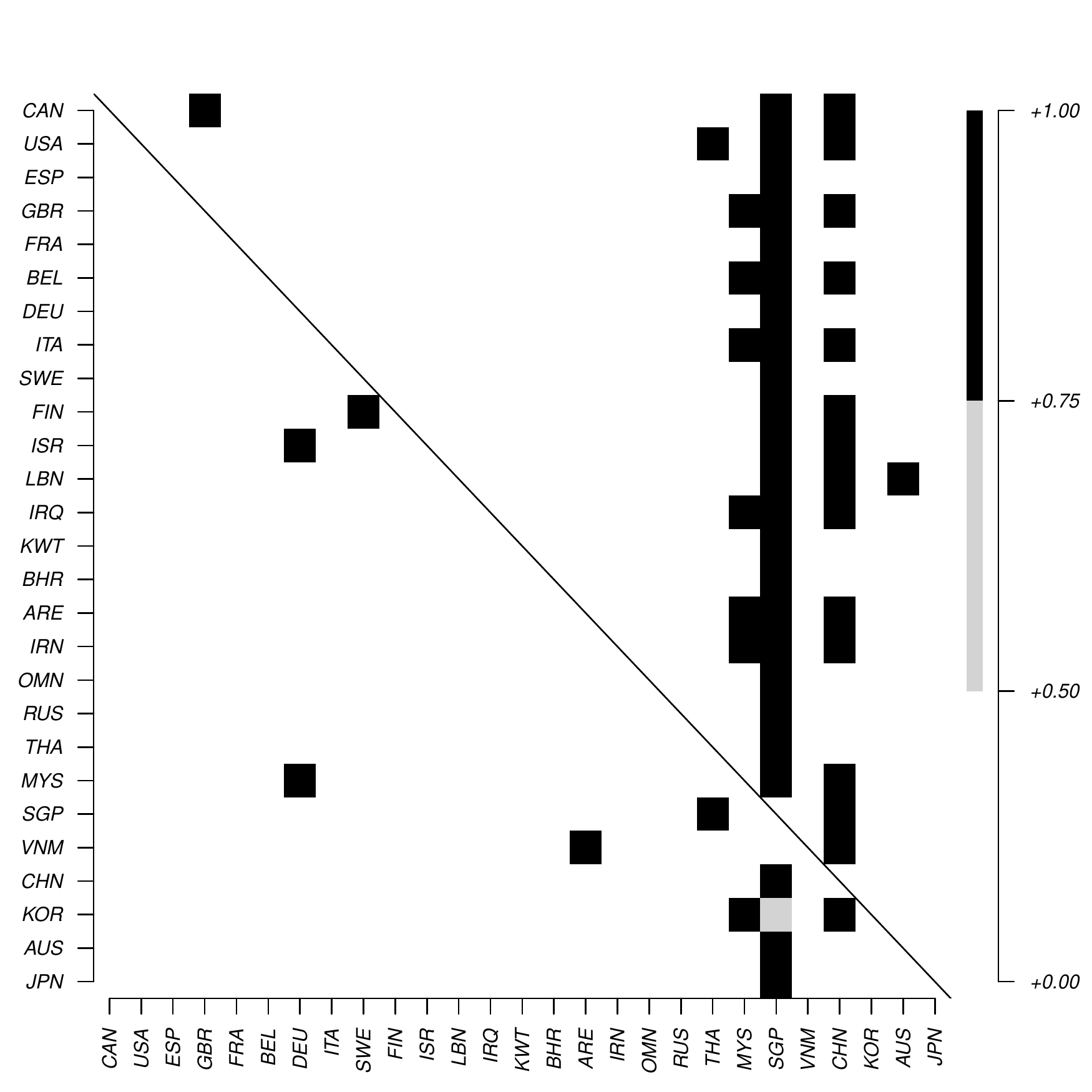}
\end{minipage}\\
\begin{minipage}{12.5cm}~\\
\scriptsize \textbf{Notes}: Posterior inclusion probabilities of spatial links based on $5,000$ MCMC draws. Inclusion probabilities 0.50-0.75 (little evidence for inclusion) are coloured grey. Strong evidence for inclusion (>0.75) indicated by black colour. 
\end{minipage}%
\end{figure}

\begin{figure}[!ht]
\caption{Posterior inclusion probabilities for specifications with $T=10$}
\label{fig:pips_spec_t10}
\centering
\begin{minipage}[b]{.45\linewidth}
\subcaption*{Fixed ($\underline{p} = 1/2$); $\boldsymbol{Wy}_t$}\vspace{-0.75cm}\label{fig:2a}
\centering \includegraphics[scale=0.35]{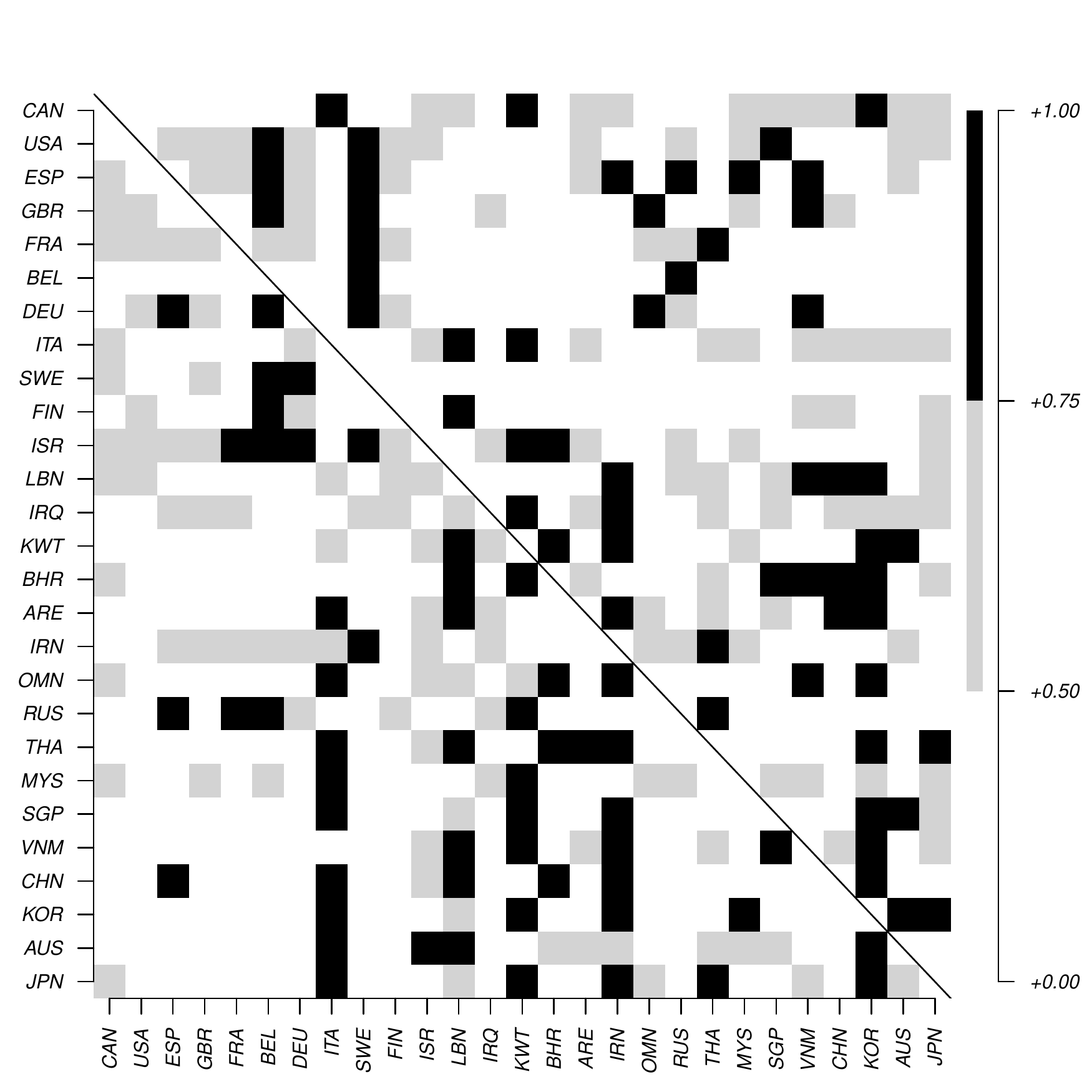}
\end{minipage}%
\begin{minipage}[b]{.45\linewidth}
\subcaption*{Sparsity ($\underline{m} = 7$); $\boldsymbol{Wy}_t$}\vspace{-0.75cm}\label{fig:2b}
\centering \includegraphics[scale=0.35]{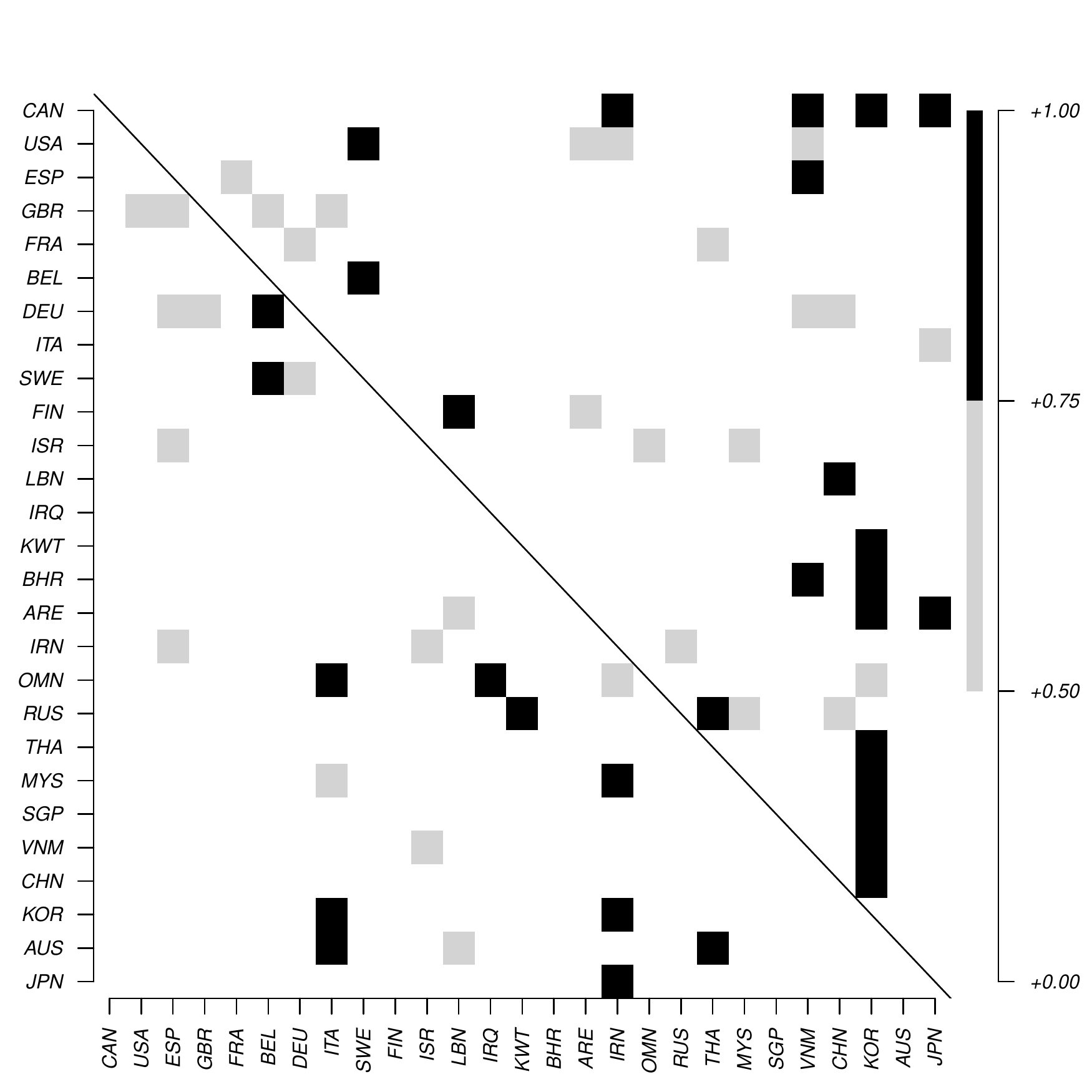}
\end{minipage}\\ \vspace{0.5cm}
\begin{minipage}[b]{.45\linewidth}
\subcaption*{Fixed ($\underline{p} = 1/2$); $\boldsymbol{Wy}_{t-14}$}\vspace{-0.75cm}\label{fig:2c}
\centering \includegraphics[scale=0.35]{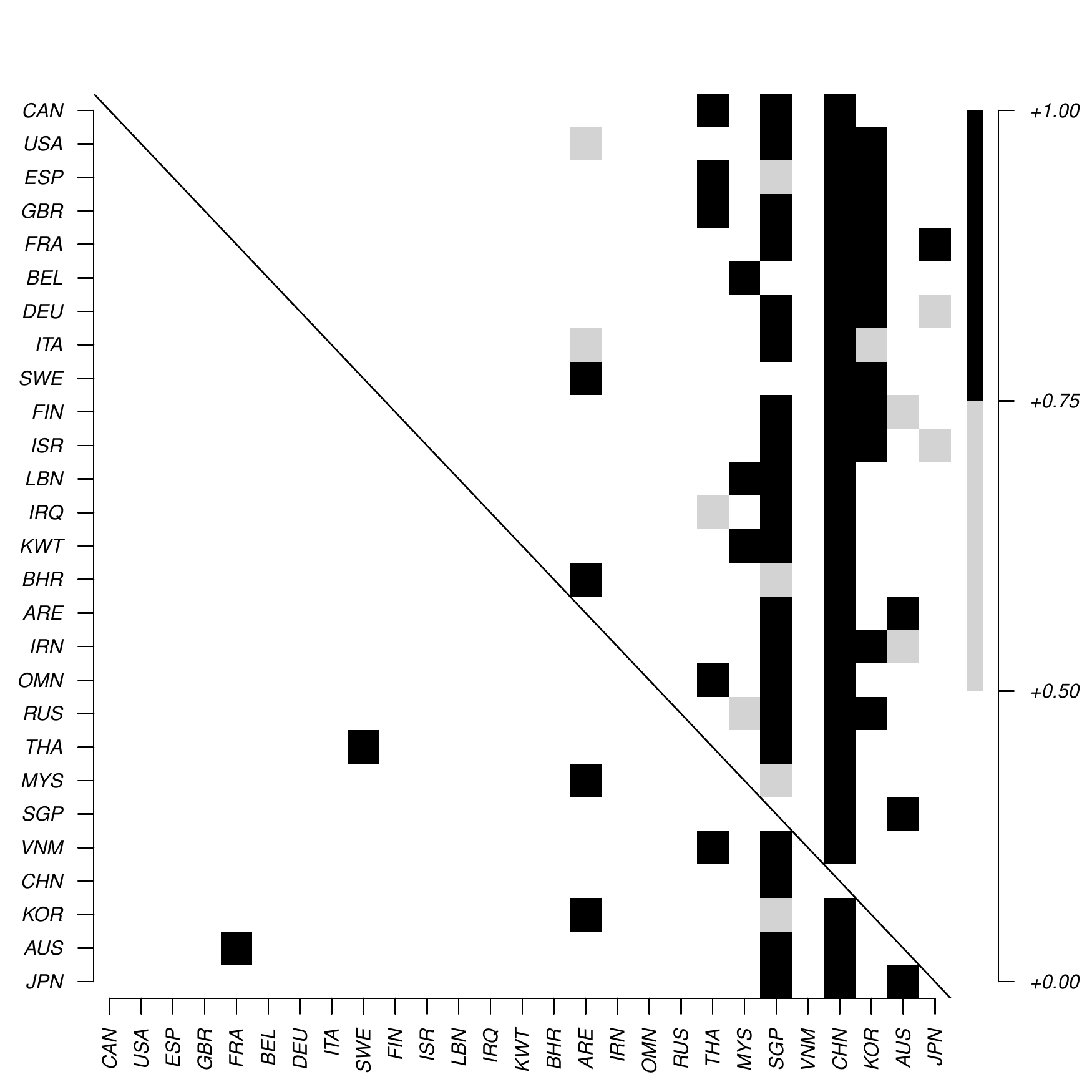}
\end{minipage}
\begin{minipage}[b]{.45\linewidth}
\subcaption*{Sparsity ($\underline{m} = 7$); $\boldsymbol{Wy}_{t-14}$}\vspace{-0.75cm}\label{fig:2d}
\centering \includegraphics[scale=0.35]{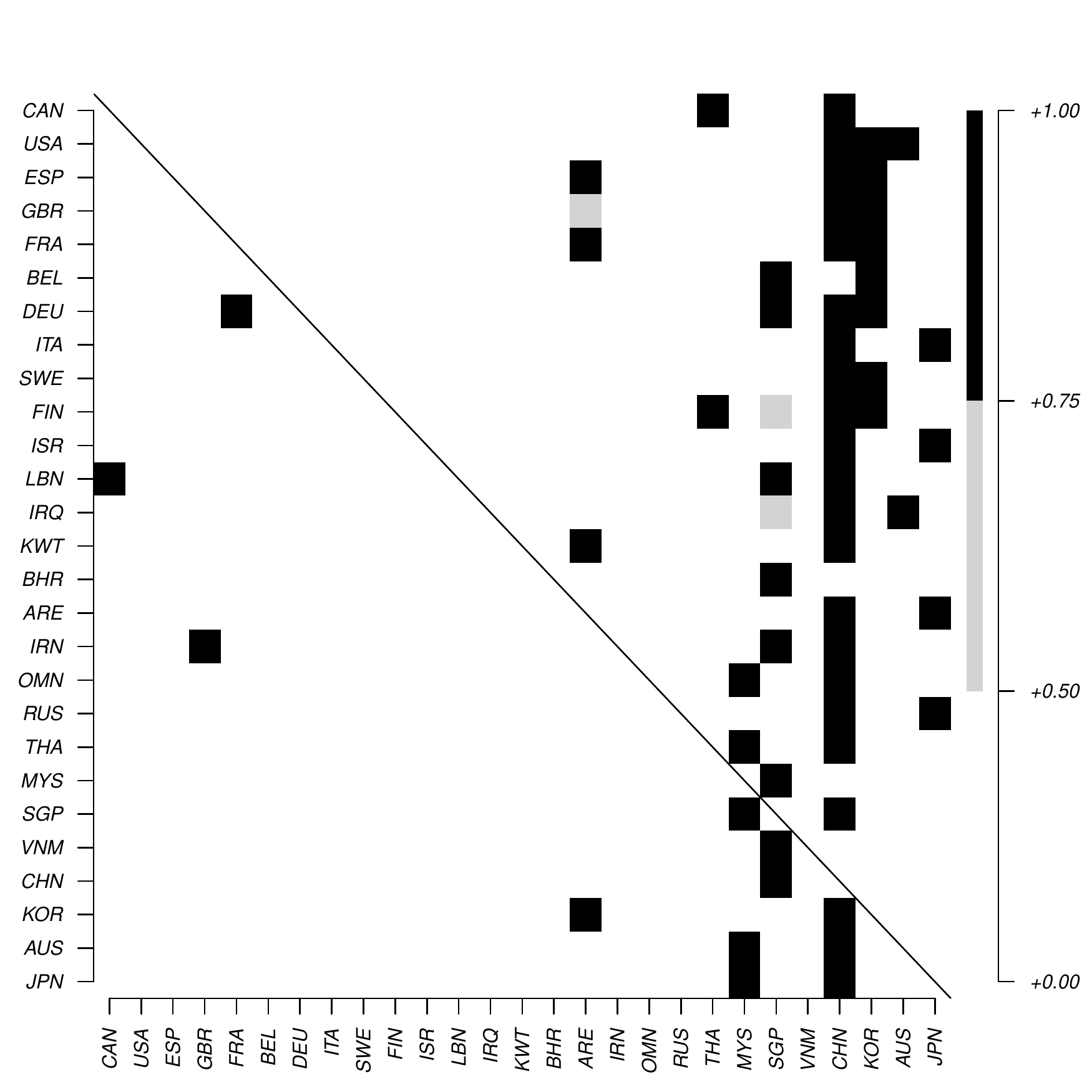}
\end{minipage}\\
\begin{minipage}{12.5cm}~\\
\scriptsize \textbf{Notes}: Posterior inclusion probabilities of spatial links based on $5,000$ MCMC draws. Inclusion probabilities 0.50-0.75 (little evidence for inclusion) are coloured grey. Strong evidence for inclusion (>0.75) indicated by black colour. 
\end{minipage}%
\end{figure}

This section provides results based on alternative model specifications. We provide estimates and inferences for three different specifications. We first consider a specification where the dependent variable is based on the log levels of infection rates rather than (biweekly) log differences (all else being equal). These results (labelled \textit{level specification} are presented in Table \ref{tab:results_level_spec}  and Figure \ref{fig:pips_level_spec}. Overall, the results in general appear very similar to the benchmark specifications.\footnote{Note that the interpretation of initial infections variable in the level specifications is different to benchmark case using log differences as dependent variable. Specifically, in the former parameters smaller than unity (as compared to negative parameters in the benchmark specifications) point towards convergence.} Second, we also consider specifications without row-standardization of the spatial weight matrix. A summary of the estimation results along with the posterior results for the spatial weigh matrix is provided in Table \ref{tab:results_notstandardized} and Figure \ref{fig:pips_notstandardized}, respectively.\footnote{From an econometric point of view, estimation is the same as compared to the row-stochastic counterparts without conducting the standardization in the MCMC sampler. However, in this case several caveats arise. Most notably, row-standardization of $\boldsymbol{W}$ has the great advantage that the parameter space for the spatial autoregressive parameter $\rho$ is clearly defined, such that the inverse $(\boldsymbol{I}_N-\rho\boldsymbol{W})^{-1}$ exists. To ensure stationarity of the MCMC sampler in case of no row-standardization, we have therefore implemented a rejection step by rejecting draws resulting to singular solutions.} Third, to show the merits of our approach in highly over-parametrized environments, we also present a robustness check with only $T=10$.\footnote{Specifically, in these specifications we reduce the end date of the dependent variable accordingly.} Results are presented in Table \ref{tab:results_t10} and Figure \ref{fig:pips_spec_t10}. We have moreover tried various other robustness checks including versions using a shorter time lag of only seven days or even shorter time periods, which produces similar results.

\begin{table}[ht]
  \centering
\begin{threeparttable}[tbp]
  \caption{Estimation results for level specifications\label{tab:results_level_spec}}
  \scriptsize
 \renewcommand{\arraystretch}{0.8}
\begin{tabular}{lrrrrrrrr}
\toprule
  & \multicolumn{4}{c}{$\boldsymbol{Wy}_{t}$} & \multicolumn{4}{c}{$\boldsymbol{Wy}_{t-14}$} \\
   & \multicolumn{2}{c}{Fixed} & \multicolumn{2}{c}{Sparsity} & \multicolumn{2}{c}{Fixed} & \multicolumn{2}{c}{Sparsity} \\
   & \multicolumn{1}{l}{Mean} & \multicolumn{1}{l}{Std.Dev.} & \multicolumn{1}{l}{Mean} & \multicolumn{1}{l}{Std.Dev.} & \multicolumn{1}{l}{Mean} & \multicolumn{1}{l}{Std.Dev.} & \multicolumn{1}{l}{Mean} & \multicolumn{1}{l}{Std.Dev.} \\
\midrule
Initial infections & \textbf{0.0438} & 0.0126 & \textbf{0.0697} & 0.0097 & \textbf{0.0222} & 0.0109 & \textbf{0.0532} & 0.0152 \\
Stringency & \textbf{-0.2633} & 0.0445 & \textbf{-0.2089} & 0.0498 & 0.0609 & 0.0398 & \textbf{-0.3677} & 0.0507 \\
Precipitation & -0.0205 & 0.0454 & -0.0077 & 0.0405 & 0.0222 & 0.0517 & 0.0075 & 0.0572 \\
Temperature & \textbf{-0.0059} & 0.0020 & \textbf{-0.0057} & 0.0018 & \textbf{-0.0032} & 0.0025 & -0.0014 & 0.0027 \\
$\rho$ & \textbf{0.9188} & 0.0139 & \textbf{0.8468} & 0.0168 & \textbf{0.9510} & 0.0155 & \textbf{0.9384} & 0.0166 \\
$\sigma^2$ & \textbf{0.0391} & 0.0029 & \textbf{0.0353} & 0.0026 & \textbf{0.0579} & 0.0044 & \textbf{0.0592} & 0.0043 \\\midrule
Avg. \# neighbours & 7.9544 &    & 3.6625 &    & 3.7096 &    & 3.1224 &  \\
Fixed effects & \text{Yes} & & \text{Yes} & & \text{Yes} & & \text{Yes} & \\ 
$N$ & 27 & & 27 & & 27 & & 27 & \\
$T$ & 19 & & 19 & & 19 & & 19 & \\
\bottomrule
\end{tabular}%
\begin{tablenotes}
\scriptsize \item \textbf{Notes:} Posterior quantities based on $5,000$ MCMC draws, where the first $2,500$ were discarded as burn-ins. Values in bold denote significance under a 90\% credible interval. Level specifications refer to specifications by using (all else being equal) log levels of infection rates rather than log-differences as the dependent variable.
\end{tablenotes}
\end{threeparttable}%
\end{table}

\begin{table}[ht]
  \centering
\begin{threeparttable}[tbp]
  \caption{Estimation results for specifications without row-standardization of the weight matrix\label{tab:results_notstandardized}}
  \scriptsize
 \renewcommand{\arraystretch}{0.8}
\begin{tabular}{lrrrrrrrr}
\toprule
  & \multicolumn{4}{c}{$\boldsymbol{Wy}_{t}$} & \multicolumn{4}{c}{$\boldsymbol{Wy}_{t-14}$} \\
   & \multicolumn{2}{c}{Fixed} & \multicolumn{2}{c}{Sparsity} & \multicolumn{2}{c}{Fixed} & \multicolumn{2}{c}{Sparsity} \\
   & \multicolumn{1}{l}{Mean} & \multicolumn{1}{l}{Std.Dev.} & \multicolumn{1}{l}{Mean} & \multicolumn{1}{l}{Std.Dev.} & \multicolumn{1}{l}{Mean} & \multicolumn{1}{l}{Std.Dev.} & \multicolumn{1}{l}{Mean} & \multicolumn{1}{l}{Std.Dev.} \\
\midrule
Initial infections & \textbf{-0.9538} & 0.0108 & \textbf{-0.9715} & 0.0093 & \textbf{-0.9718} & 0.0217 & \textbf{-0.9655} & 0.0223 \\
Stringency & \textbf{-0.4062} & 0.0472 & \textbf{-0.5259} & 0.0360 & \textbf{-0.4856} & 0.0739 & \textbf{-0.3712} & 0.0738 \\
Precipitation & 0.0084 & 0.0365 & -0.0304 & 0.0357 & -0.0966 & 0.1023 & -0.1284 & 0.1071 \\
Temperature & \textbf{-0.0025} & 0.0017 & -0.0018 & 0.0017 & \textbf{-0.0116} & 0.0047 & \textbf{-0.0172} & 0.0048 \\
$\rho$ & \textbf{0.0801} & 0.0018 & \textbf{0.0960} & 0.0026 & \textbf{0.3624} & 0.0452 & \textbf{0.2526} & 0.0484 \\
$\sigma^2$ & \textbf{0.0264} & 0.0018 & \textbf{0.0266} & 0.0018 & \textbf{0.2318} & 0.0151 & \textbf{0.2457} & 0.0164 \\\midrule
Avg. \# neighbours & 10.8863 &    & 6.5014 &    & 2.5103 &    & 2.0225 &  \\
Fixed effects & \text{Yes} & & \text{Yes} & & \text{Yes} & & \text{Yes} & \\ 
$N$ & 27 & & 27 & & 27 & & 27 & \\
$T$ & 19 & & 19 & & 19 & & 19 & \\
\bottomrule
\end{tabular}%
\begin{tablenotes}
\scriptsize \item \textbf{Notes:} Posterior quantities based on $5,000$ MCMC draws, where the first $2,500$ were discarded as burn-ins. Values in bold denote significance under a 90\% credible interval.
\end{tablenotes}
\end{threeparttable}%
\end{table}

\begin{table}[ht]
  \centering
\begin{threeparttable}[tbp]
  \caption{Estimation results for specifications with $T=10$\label{tab:results_t10}}
  \scriptsize
 \renewcommand{\arraystretch}{0.8}
\begin{tabular}{lrrrrrrrr}
\toprule
  & \multicolumn{4}{c}{$\boldsymbol{Wy}_{t}$} & \multicolumn{4}{c}{$\boldsymbol{Wy}_{t-14}$} \\
   & \multicolumn{2}{c}{Fixed} & \multicolumn{2}{c}{Sparsity} & \multicolumn{2}{c}{Fixed} & \multicolumn{2}{c}{Sparsity} \\
   & \multicolumn{1}{l}{Mean} & \multicolumn{1}{l}{Std.Dev.} & \multicolumn{1}{l}{Mean} & \multicolumn{1}{l}{Std.Dev.} & \multicolumn{1}{l}{Mean} & \multicolumn{1}{l}{Std.Dev.} & \multicolumn{1}{l}{Mean} & \multicolumn{1}{l}{Std.Dev.} \\
\midrule
Initial infections & \textbf{-0.9834} & 0.0161 & \textbf{-1.0090} & 0.0122 & \textbf{-1.0234} & 0.0207 & \textbf{-1.0016} & 0.0175 \\
Stringency & \textbf{-0.2595} & 0.0853 & \textbf{-0.4966} & 0.0672 & -0.1537 & 0.1667 & 0.0200 & 0.0822 \\
Precipitation & -0.0067 & 0.0453 & 0.0131 & 0.0440 & -0.1114 & 0.0764 & -0.1084 & 0.0724 \\
Temperature & 0.0030 & 0.0022 & -0.0022 & 0.0024 & \textbf{-0.0093} & 0.0038 & \textbf{-0.0062} & 0.0037 \\
$\rho$ & \textbf{0.7148} & 0.0111 & \textbf{0.4405} & 0.0164 & \textbf{0.8910} & 0.0317 & \textbf{0.8930} & 0.0359 \\
$\sigma^2$ & \textbf{0.0183} & 0.0019 & \textbf{0.0193} & 0.0019 & \textbf{0.0570} & 0.0060 & \textbf{0.0513} & 0.0064 \\\midrule
Avg. \# neighbours & 10.4409 &    & 3.5630 &    & 3.1953 &    & 2.4116 &  \\
Fixed effects & \text{Yes} & & \text{Yes} & & \text{Yes} & & \text{Yes} & \\ 
$N$ & 27 & & 27 & & 27 & & 27 & \\
$T$ & 10 & & 10 & & 10 & & 10 & \\
\bottomrule
\end{tabular}%
\begin{tablenotes}
\scriptsize \item \textbf{Notes:} Posterior quantities based on $5,000$ MCMC draws, where the first $2,500$ were discarded as burn-ins. Values in bold denote significance under a 90\% credible interval.
\end{tablenotes}
\end{threeparttable}%
\end{table}

\begin{figure}[!ht]
\caption{Diagnostic plots for a Monte Carlo run based on $N=20$ and $T=10$}
\label{fig:traceplot_MCMC}
\centering
\begin{minipage}[b]{.5\linewidth}
\subcaption*{Fixed ($\underline{p} = 1/2$)\\ $\tilde{\rho} = 0.3$}
\centering \includegraphics[scale=0.4]{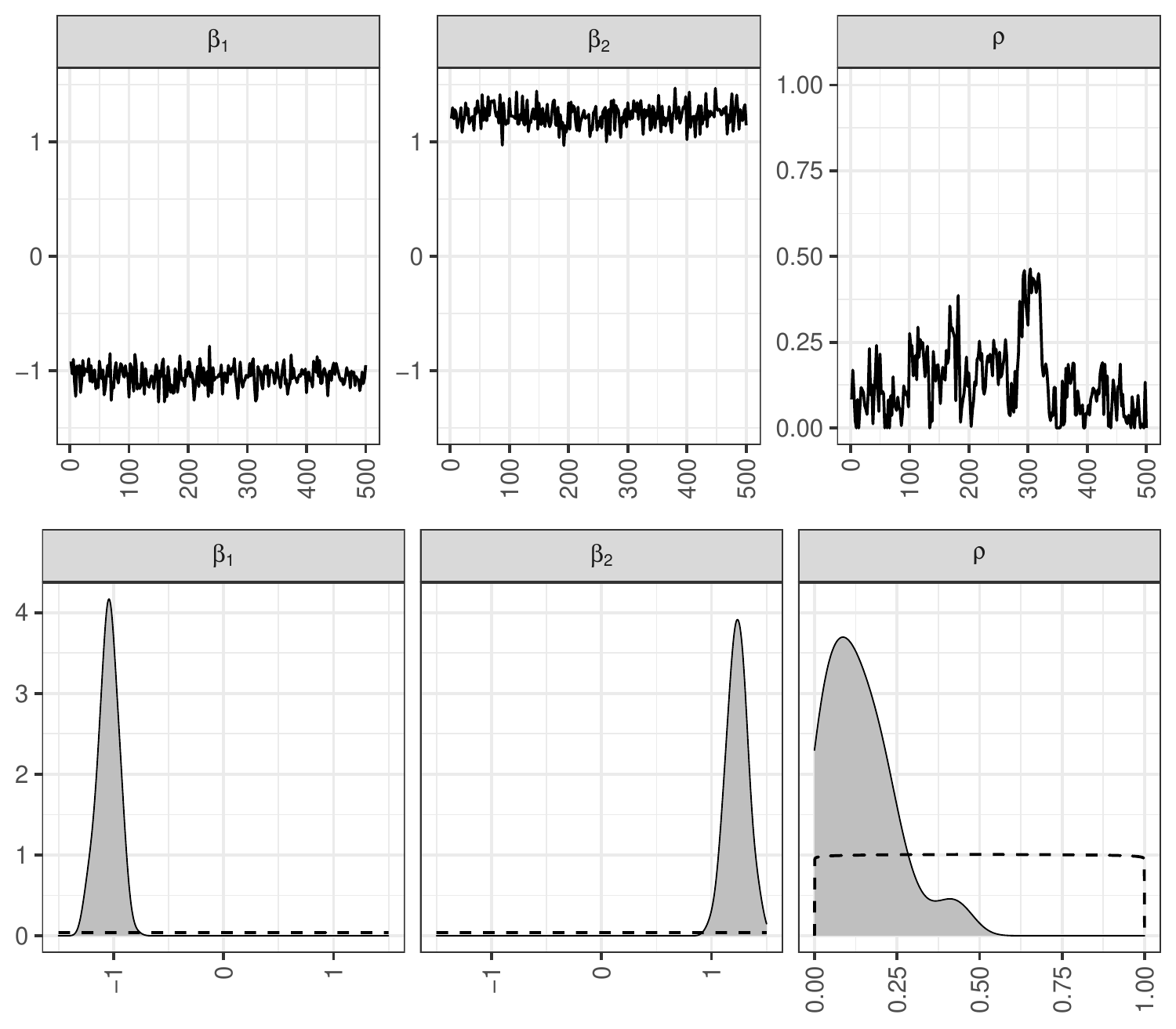}
\end{minipage}%
\begin{minipage}[b]{.5\linewidth}
\subcaption*{Sparsity ($\underline{m} = N/10$)\\ $\tilde{\rho} = 0.3$}
\centering \includegraphics[scale=0.4]{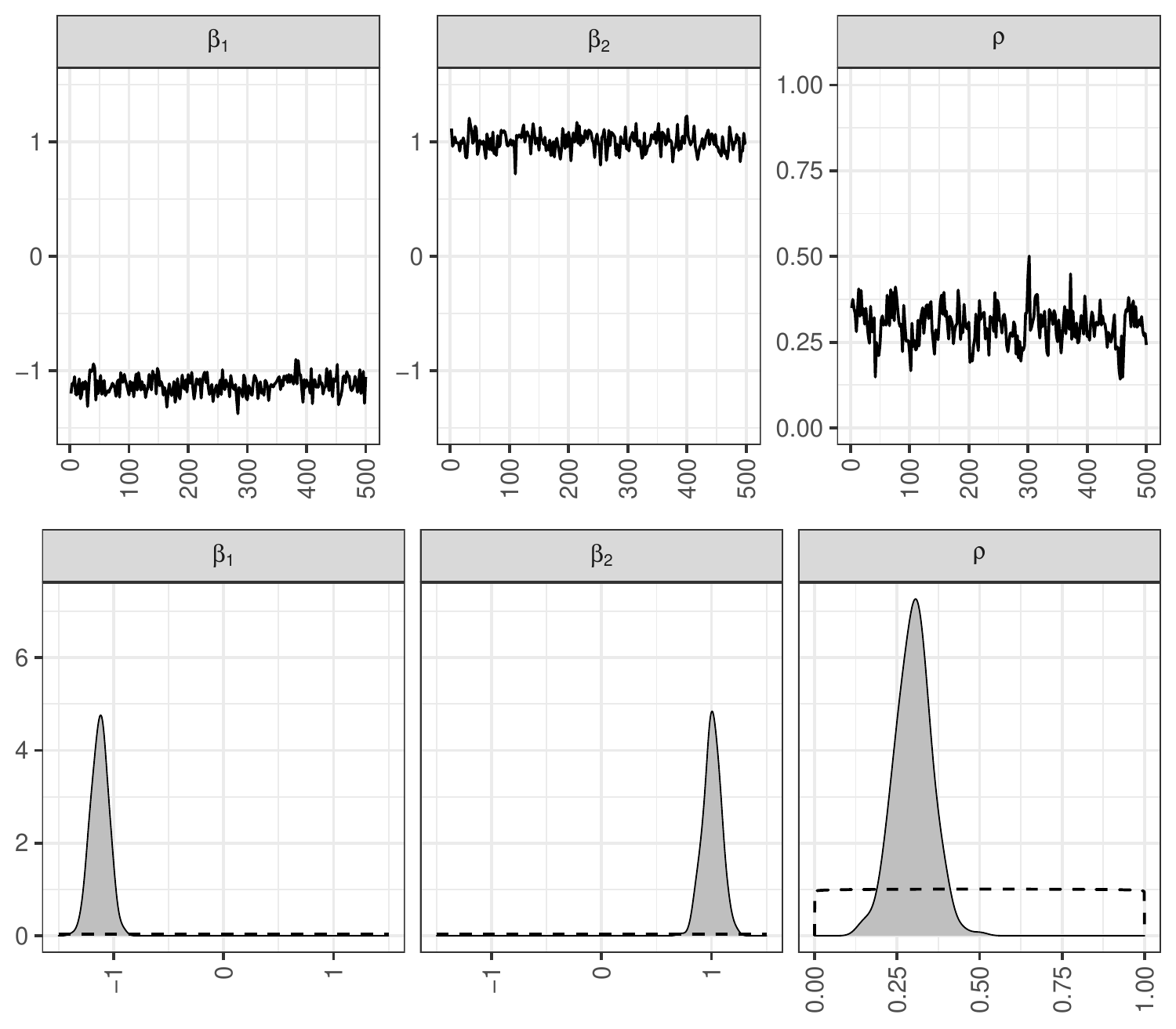}
\end{minipage}\\ 
\begin{minipage}[b]{.5\linewidth}
\subcaption*{$\tilde{\rho} = 0.5$}
\centering \includegraphics[scale=0.4]{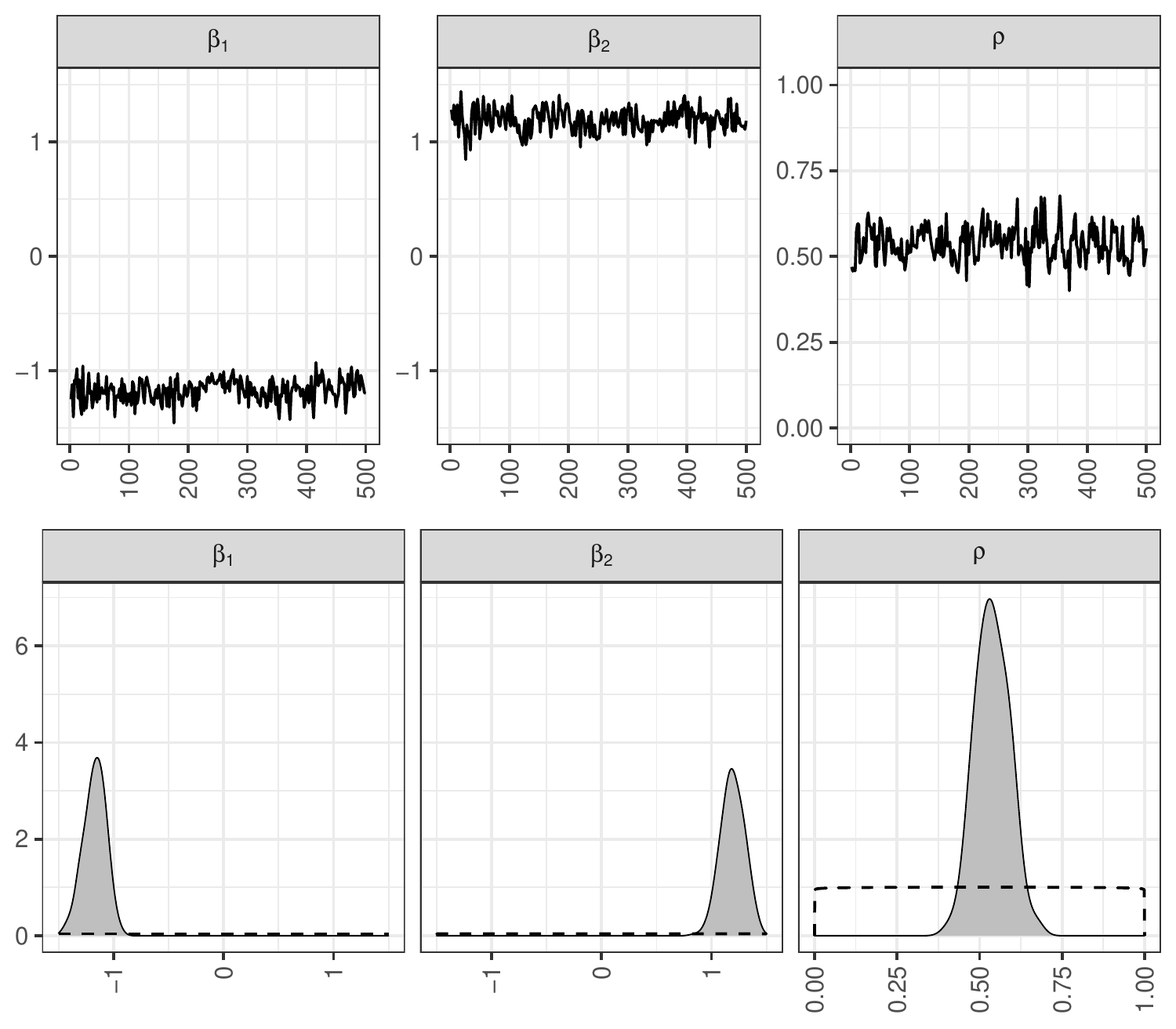}
\end{minipage}%
\begin{minipage}[b]{.5\linewidth}
\subcaption*{$\tilde{\rho} = 0.5$}
\centering \includegraphics[scale=0.4]{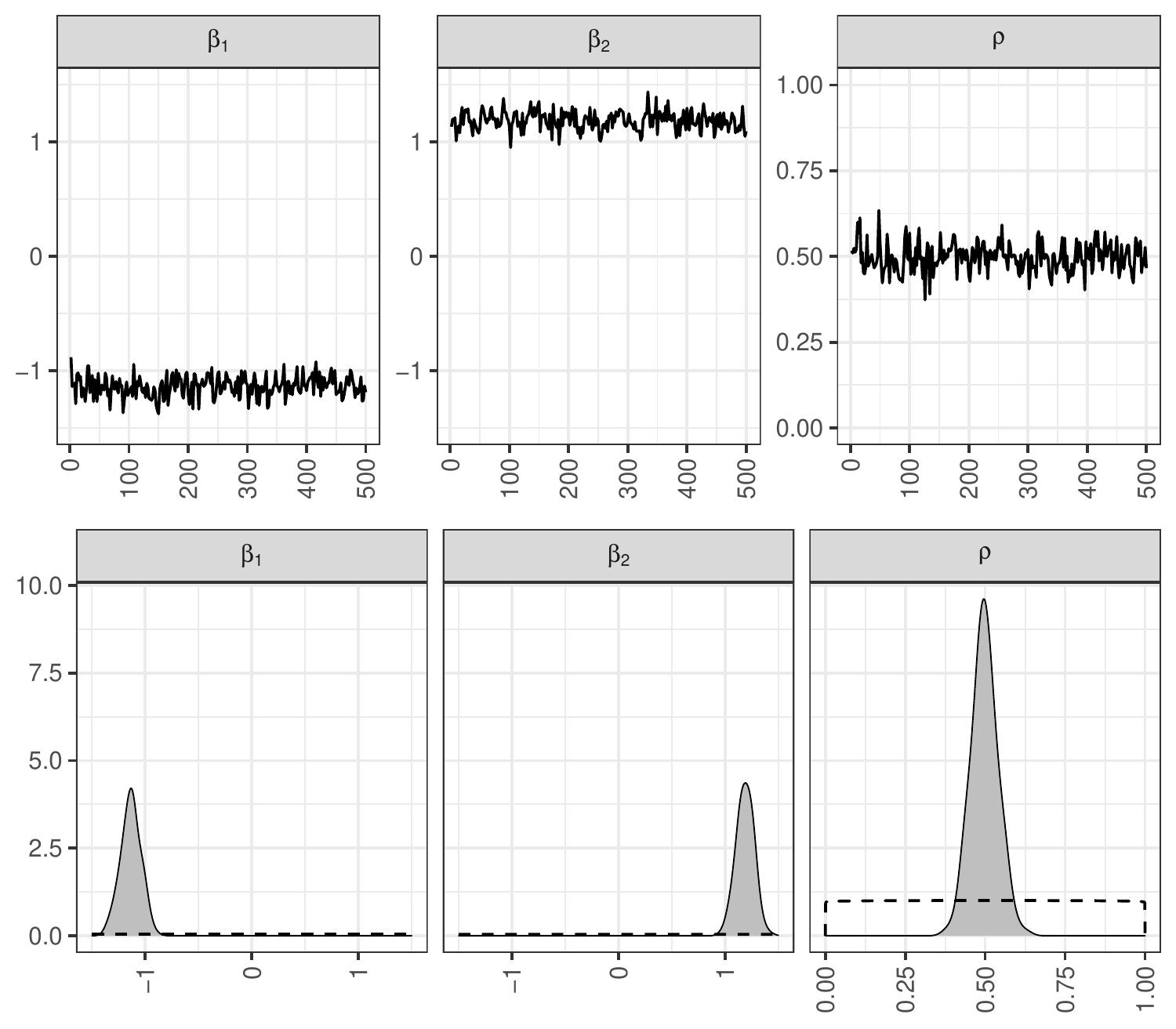}
\end{minipage}\\ 
\begin{minipage}[b]{.5\linewidth}
\subcaption*{$\tilde{\rho} = 0.8$}
\centering \includegraphics[scale=0.4]{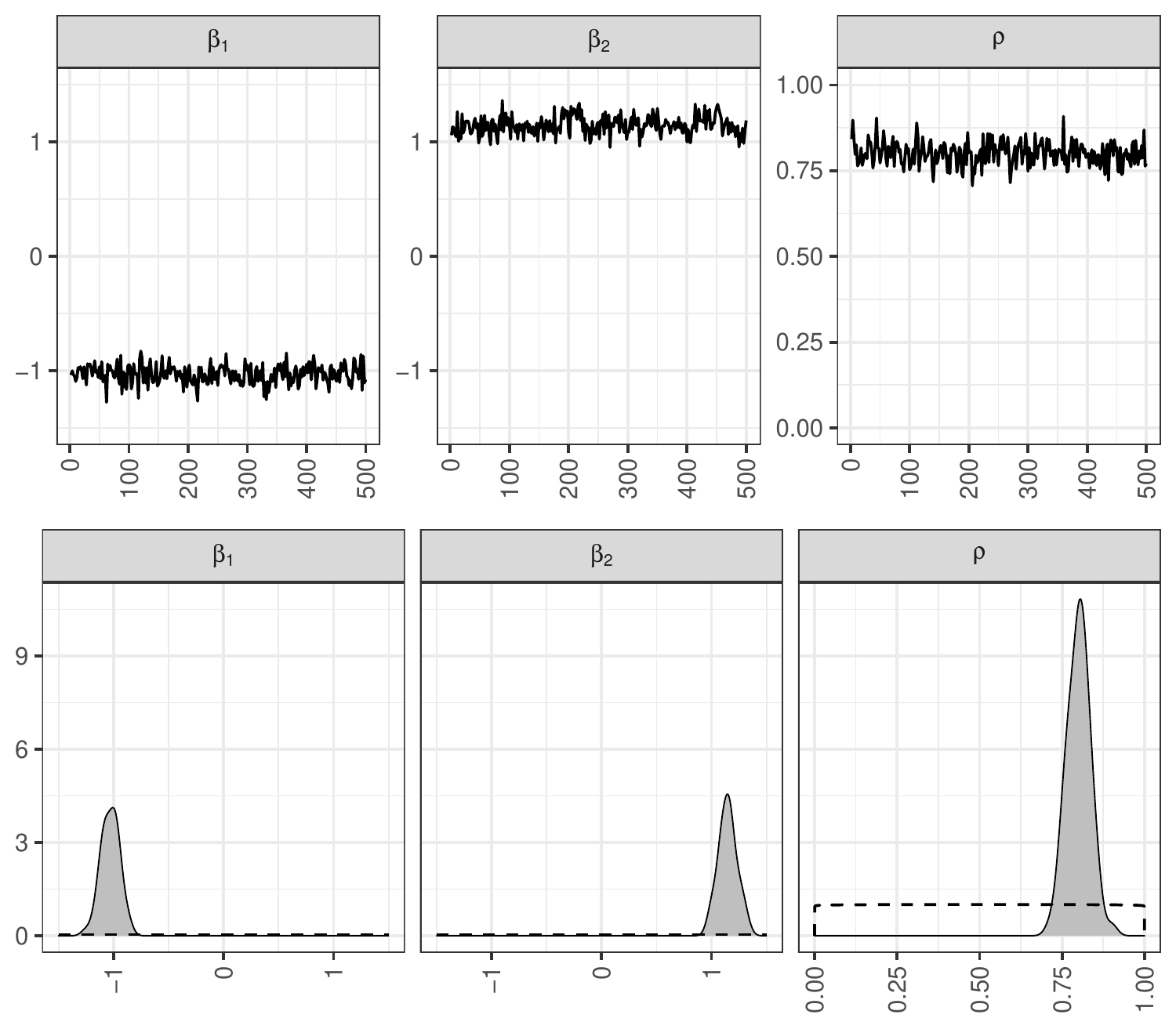}
\end{minipage}%
\begin{minipage}[b]{.5\linewidth}
\subcaption*{$\tilde{\rho} = 0.8$}
\centering \includegraphics[scale=0.4]{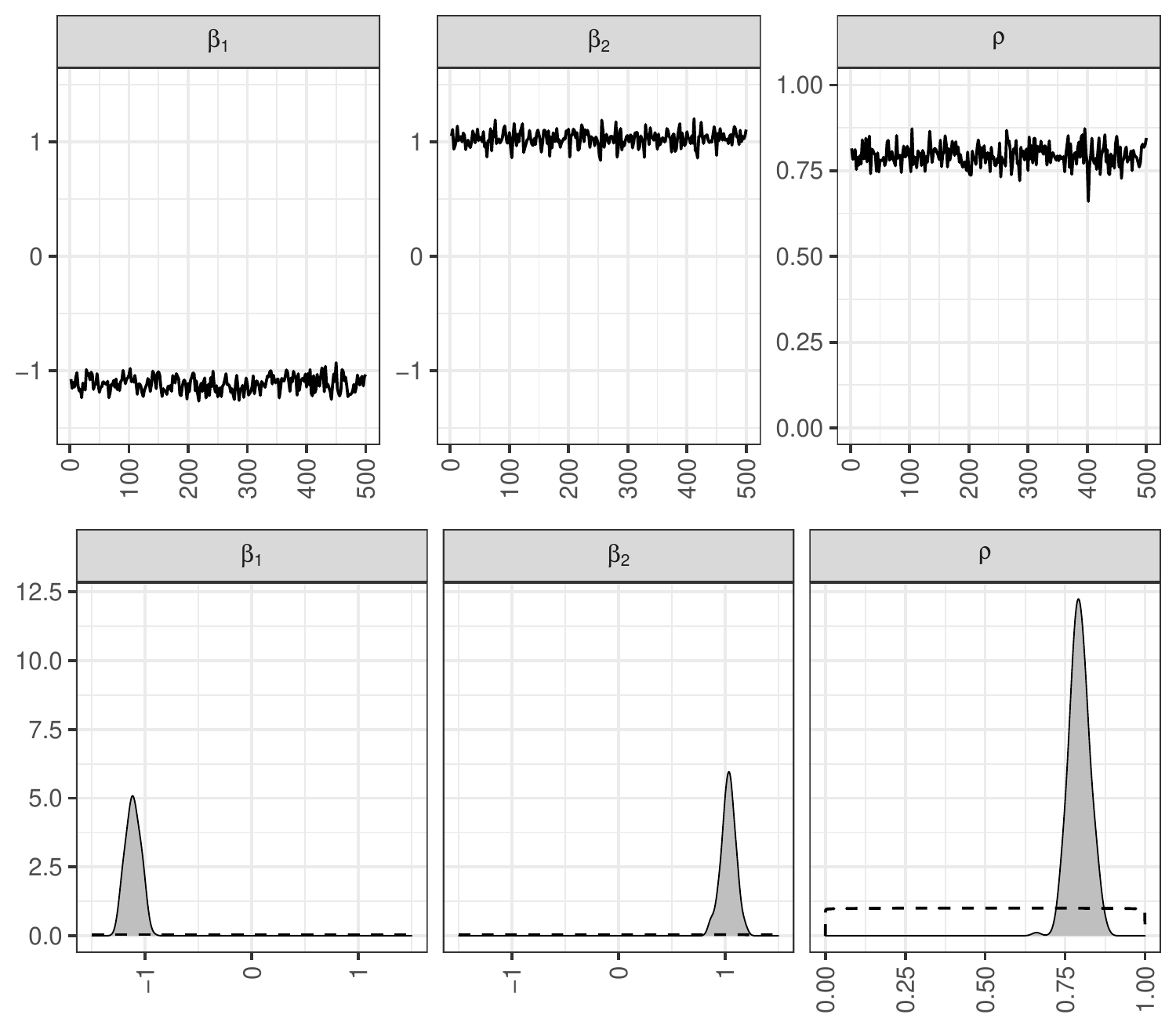}
\end{minipage}\\ 
\begin{minipage}{\linewidth}~\\
\scriptsize \textbf{Notes}: Trace plots and posterior densities based on $1,000$ MCMC draws, where the first $500$ were discarded as burn-ins. Dashed lines denote prior distributions.
\end{minipage}%
\end{figure}

\end{document}